\documentclass[fleqn,usenatbib]{mnras}
\pdfoutput=1
\usepackage{newtxtext,newtxmath}
\usepackage[T1]{fontenc}
\usepackage{ae,aecompl}
\usepackage{natbib,textcomp,placeins,epsfig,graphicx,color,amsmath,amssymb,upgreek,listings,epstopdf}

\newcommand{\teff}{$T_{\mbox{\footnotesize eff}}$}
 
 \newcommand{\kpc}{\,{\rm kpc}}
\newcommand{\kms}{\,{\rm km}\,{\rm s}^{-1}}
 
\newcommand{\Rgal}{$R_{\rm Gal}$}
\newcommand{\Gaia}{{\it Gaia}\,}
\newcommand{\GES}{{\it Gaia}-Enceladus-Sausage\,}

\title[SkyMapper view of the \GES]{The SkyMapper-\Gaia RVS view of the \GES -- an investigation of the metallicity and mass of the Milky Way's last major merger}

\author[D. K. Feuillet et al.]{Diane K. Feuillet,$^{1}$\thanks{email: feuillet@astro.lu.se} 
Sofia Feltzing,$^{1}$
Christian L. Sahlholdt,$^{1}$
Luca Casagrande$^{2,3}$
\\
$^{1}$Lund Observatory, Department of Astronomy and Theoretical Physics, Box 43, SE-221\,00 Lund, Sweden\\
$^{2}$Research School of Astronomy and Astrophysics, Mount Stromlo Observatory, The Australian National University, ACT 2611, Australia\\
$^{3}$ARC Centre of Excellence for All Sky Astrophysics in 3 Dimensions (ASTRO 3D)\\
}

\date{Accepted XXX. Received YYY; in original form ZZZ}

\pubyear{2019}

\begin{document}
\label{firstpage}
\pagerange{\pageref{firstpage}--\pageref{lastpage}}
\maketitle

\begin{abstract}

We characterize the \GES kinematic structure recently discovered in the Galactic halo using photometric metallicities from the SkyMapper survey, and kinematics from \Gaia radial velocities measurements. By examining the metallicity distribution functions (MDFs) of stars binned in kinematic/action spaces, we find that the $\sqrt{J_R}$ vs $L_z$ space allows for the cleanest selection of \GES stars with minimal contamination from disc or halo stars formed {\it in situ} or in other past mergers. Stars with $30 \leq \sqrt{J_R} \leq 50$ (kpc km s$^{-1})^{1/2}$ and $-500 \leq L_z \leq 500$ kpc km s$^{-1}$ have a narrow MDF centered at [Fe/H] $= -1.17$ dex with a dispersion of 0.34 dex. This [Fe/H] estimate is more metal-rich than literature estimates by $0.1-0.3$ dex. Based on the MDFs, we find that selection of \GES stars in other kinematic/action spaces without additional population information leads to contaminated samples. The clean \GES sample selected according to our criteria is {\it slightly} retrograde and lies along the blue sequence of the high $V_T$ halo CMD dual sequence. Using a galaxy mass-metallicity relation derived from cosmological simulations and assuming a mean stellar age of 10 Gyr we estimate the mass of the \GES progenitor satellite to be $10^{8.85-9.85}$ M$_{\odot}$, which is consistent with literature estimates based on disc dynamic and simulations. Additional information on detailed abundances and ages would be needed for a more sophisticated selection of purely \GES stars.

\end{abstract}


\section{Introduction}

The Milky Way halo contains evidence of past mergers in the form of stellar streams \citep[e.g.][]{Belokurov2006}. These overdensities of stars on the sky move together and are the remnants of dwarf galaxies that accreted onto our Galaxy, contributing to the stellar halo \citep[e.g.][]{HelmiWhite1999}. However, the origins of the diffuse stellar halo population remained unclear. Many observational studies have advocated for a dual nature of the halo \citep[e.g.][]{Carollo2007, Nissen2010}, suggesting perhaps there exists a major accreted population \citep[e.g.][]{Hayes2018} that is spatially integrated into the {\it in situ} classical halo \citep[see][]{Eggen1962}. 
Most cosmological simulations produce Milky Way-like galaxies with halos containing a significant accreted component \citep[e.g.][]{Cooper2010}.

\Gaia Data Release 2 \citep[DR2,][]{Gaia2018a} has greatly expanded our view of the Milky Way by providing precise astrometry to millions of stars. Significant substructures, beyond the visual overdensities of the stellar streams, have been identified in kinematic space which suggest at least one major merger event contributed to building the Milky Way halo. The largest kinematic structures recently found are the {\it Gaia}-Sausage \citep[][]{Belokurov2018} and {\it Gaia}-Enceladus \citep[][]{Helmi2018}. The extend to which these structures represent the same accretion event is still unclear, however, there are some differences in their initial identification. The {\it Gaia}-Sausage was first identified by \citet{Belokurov2018} in the $V_{\phi}$ vs $V_R$ velocity space as a `Sausage-like' structure centered around $V_{\phi} \sim 0$ and extended in $V_R$. Using a suite of cosmological simulations they estimate that the mass of the progenitor must have been $M_{\footnotesize \mbox{vir}} > 10^{10}$ M$_{\odot}$.
{\it Gaia}-Enceladus was first identified by \citet{Helmi2018} as stars in the inner halo with kinematics ranging from highly eccentric to highly retrograde, selected in $L_z$ and $E_n$. Using APOGEE data, \citet{Helmi2018} find that these stars lie along an [$\alpha$/Fe] vs [Fe/H] track that is more metal-poor than the thin disc and lower in [$\alpha$/Fe] than the thick disc. \citet{Helmi2018} find {\it Gaia}-Enceladus to have a mean [Fe/H] $\sim -1.6$ from APOGEE data and constrain the age to be 10-13 Gyr. They argue that a progenitor stellar mass of $6 \times 10^8$ M$_{\odot}$ is consistent with the chemical evolution sequence shown in the APOGEE data.

As there still are some discussion in the literature about the exact nature and definition of the old merger event, we will call it \GES in order to acknowledge the current uncertainty and the growing concensus in the community about its name. In the following we will thus refer to the \GES when we talk about the merged galaxy in general. When we discuss the specific selection done in a paper we will refer to the structure by the name used in that study.

There have also been kinematic substructures identified based on the orbital properties, age, and metallicity of globular clusters such as the Sequoia event identified by \citet{Myeong2019} based on 10 high-energy, high-eccentricity globular clusters \citep{Myeong2018} and the {\it Kraken} \citep{Kruijssen2019}. \citet{Myeong2019} argue that the Sequoia and {\it Gaia}-Sausage are distinct accretion events that make up {\it Gaia}-Enceladus has identified by \citet{Helmi2018}. The {\it Gaia}-Sausage stars have high-energy, radial orbits and a mean [Fe/H] $\sim -1.3$ while the Sequoia stars are highly retrograde with a mean [Fe/H] $\sim -1.6$. \citet{Koppelman2019} use a clustering algorithm to identify several kinematic substructures in the high velocity stars. The main structures they identify are {\it Gaia}-Enceladus, the Sequoia, the Helmi Streams \citep{Helmi1999}, and two new structures with low-energy, retrograde orbits that they dub Thamnos 1 and 2. The {\it Gaia}-Enceladus structure selected with the clustering algorithm of \citet{Koppelman2019} is kinematically more confined than the \citet{Helmi2018} selection. In their Figure~2, the region of the Toomre diagram occupied by {\it Gaia}-Enceladus stars in \citet{Helmi2018} is now assigned to four different substructures. The revised {\it Gaia}-Enceladus selection by \citet{Koppelman2019} results in stars constrained to large $V_R$ with kinematics consistent with the \citet{Belokurov2018} {\it Gaia}-Sausage. We hereafter refer to this kinematic structure in the Milky Way as the \GES and use the \citet{Belokurov2018} and \citet{Koppelman2019} structures to guide our search.

\Gaia DR2 also revealed the existence of a double sequence in the Hertzsprung-Russell Diagram (HRD) of the high velocity halo \citep{Gaia2018b}, which could be associated with the two halo groups noted by \citet{Nissen2010}. The two sequences are easily visible in both the dwarfs and the giants, but are more often studied using the giants, where they are designated as the red and blue sequences. \citet{Haywood2018} find that while the two sequences overlap in kinematic space, it is likely that the blue sequence is composed of accreted stars and the red sequence is composed of kinematically heated Milky Way stars, either the old thick disc \citep[][]{DiMatteo2019}, {\it `in situ'} halo \citep[e.g.][]{Zolotov2009}, or a mix \citep{Amarante2020a}. \citet{Gallart2019} have similar conclusions, finding the age distribution of the red and blue sequences are identically old, cutting off sharply at 10 Gyr. They associate this cutoff, which is immediately followed by a peak in thick disc stars, with the infall of the \GES progenitor. \citet{Sahlholdt2019} characterize the giants of the dual sequences using SkyMapper data, finding that the blue sequence has a peak metallicity of $-1.4$ dex while the red sequence has a peak metallicity of $-0.7$ dex, in agreement with the [Fe/H] distributions of \citet{Gallart2019}. \citet{Sahlholdt2019} also find the two sequences have the same age.

\begin{figure*}
\includegraphics[clip,width=0.99\hsize,angle=0,trim=2cm 0cm 0.9cm 0cm]{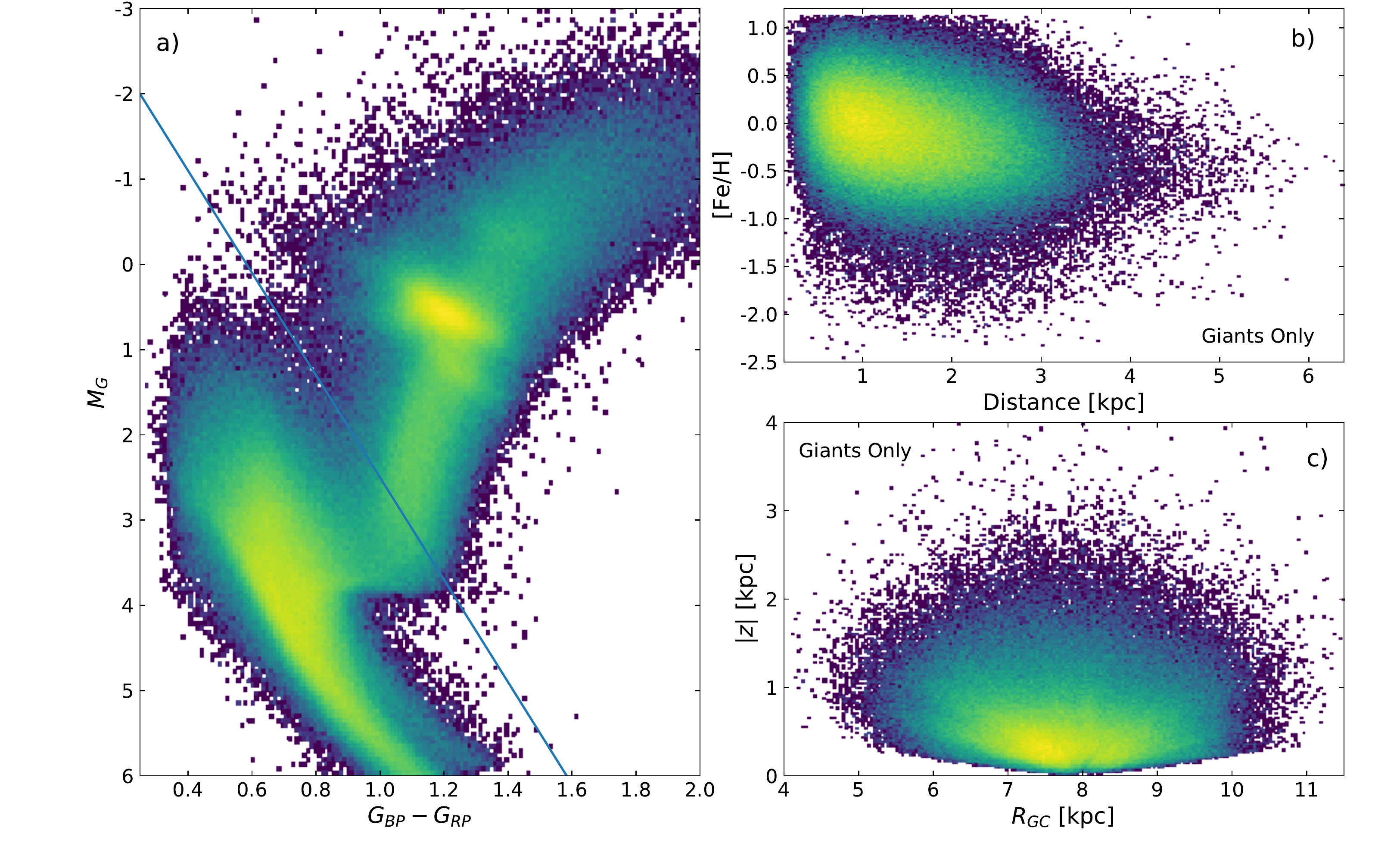}
\caption{a) The log(density) CMD of the SkyMapper-RVS sample, described in section \ref{sec:data}. The giant selection is indicated by the blue line; stars brighter than this line are taken as giants. b) The [Fe/H] vs distance log(density) distribution of the giant sample. c) The $|z|$ vs \Rgal\, log(density) distribution of the giant sample. }
\label{fig:CMD_all}
\end{figure*}

Most of the kinematic studies have had limited information on the metallicity of the structures. [Fe/H] estimates have been made using small numbers of stars cross-matched with spectroscopic surveys \citep[e.g.][]{Helmi2018, Myeong2019} or using color-magnitude fitting to a population of stars \citep{Gallart2019}. However, to fully characterize a population, metallicities of individual stars in higher numbers are needed to avoid contamination and provide full metallicity distribution functions (MDFs). Another hurdle in studying these kinematic substructures is selecting clean populations. Samples of accreted stars selected using a large region of kinematic space can easily be contaminated by heated disc or {\it `in situ'} halo stars. In this paper, we use the SkyMapper photometric metallicities determined by \citet{Casagrande2019}, which includes 10 million \Gaia DR2 stars, to explore the metallicity variations over the kinematic space containing possible \GES stars. 

\section{Data}
\label{sec:data}

The sample presented is a cross match of the SkyMapper Southern Sky Survey \citep{Casagrande2019} with the \Gaia DR2 Radial Velocity Spectrometer (RVS) catalogue \citep{Gaia2018a}, resulting in $\sim 900,000$ stars with photometric \teff, photometric [Fe/H], and full 6 dimensional phase-space coordinates. The sample is limited to parallax uncertainties $< 10\%$ to ensure minimal uncertainties in the derived kinematics from radial velocity, proper motion, and distance errors. Using \Gaia RVS limits the sample to $G$-band magnitudes $\lesssim 13$, which limits the volume ($< 3$ kpc for red clump and $< 10$ kpc for luminous giants, assuming no extinction) but allows for full kinematic measurements. The present sample reaches $\sim 3.5\,\kpc$ from the Sun, see Figure~\ref{fig:CMD_all} panel~b.

We use the distance estimates provided by \citet{Schonrich2019} using a parallax offset of 0.054 mas and a parallax error that is increased by 0.043~mas. The \citet{Schonrich2019} catalogue accounts for the full parallax offset, beyond the offset originally reported by the \Gaia team at the time of DR2. Although distance catalogues using more complex distance estimation techniques, such as StarHorse \citep{Queiroz2018}, are available for our sample, we feel a simpler distance estimation method is reliable for such a nearby sample. 

The extinction and reddening are derived using the $E(B-V)$ provided by SkyMapper \citep{Casagrande2019} and the coefficients provided by \citet{Casagrande2018} to convert to \Gaia magnitudes and colors. The de-reddened color magnitude diagram (CMD) of the full SkyMapper-RVS sample is shown in Figure~\ref{fig:CMD_all} panel~a. In this sample features such as the binary sequence, red giant branch bump, asymptotic giant branch bump, and red clump are clearly visible. 

The photometric metallicities of the SkyMapper survey are known to have offsets for the metal-poor subgiant stars. As we are interested in characterizing the metallicity distribution of the sample in question, we limit our study to the giant stars. The blue line in Figure~\ref{fig:CMD_all} panel a shows the giant selection. Only stars brighter than this line are used in this study, resulting in a sample of 372,000 stars. The mean [Fe/H] uncertainty for the giant sample is 0.17~dex. 
Figure~\ref{fig:CMD_all} also shows the [Fe/H] vs distance distribution (panel b) and the $|z|$ vs \Rgal\, distribution (panel c) for the giants only. As mentioned above, the $G \lesssim 13$ magnitude limit of \Gaia RVS results in a sample reaching $\sim 3.5 \kpc$ from the Sun. 

\begin{figure*}
\includegraphics[clip,width=0.99\hsize,angle=0,trim=2.5cm 3cm 3.5cm 3cm] {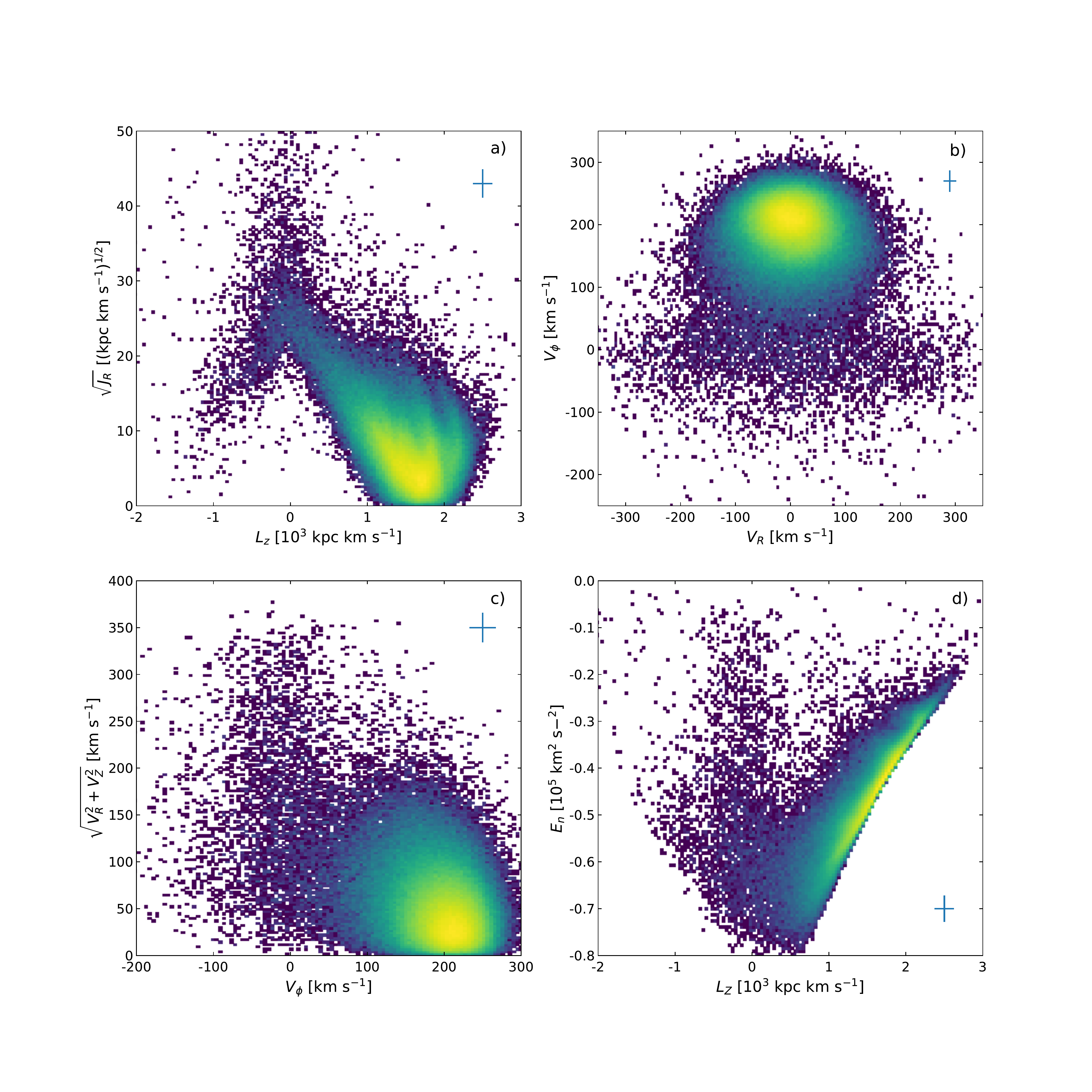}
\caption{The log(density) kinematic distributions of the full SkyMapper-RVS giant sample. a) $\sqrt{J_R}$ vs $L_z$, b) $V_{\phi}$ vs $V_R$, c) $\sqrt{V_R^2 + V_Z^2}$ vs $V_{\phi}$, and d) $E_n$ vs $L_z$. The mean uncertainty is given by the blue error bar in each panel. }
\label{fig:full_kin}
\end{figure*}

Full space velocities, actions, and orbits energy were calculated for the giant sample with {\it galpy} using the `MWPotential2014' potential \citep{Bovy2015}. Full orbit integrations were only performed for a representative sample of \GES stars, see Appendix \ref{ap:orbits}. Figure~\ref{fig:full_kin} shows the kinematic distributions for the giant sample in four phase spaces typically explored in the literature: a) $\sqrt{J_R}$ vs $L_z$, b) $V_{\phi}$ vs $V_R$, c) $\sqrt{V_R^2 + V_Z^2}$ vs $V_{\phi}$, and d) $E_n$ vs $L_z$. In $\sqrt{J_R}$ vs $L_z$ the disc substructure can be seen between $L_z$ of 1.0 and 2.5, as discussed by \citet{Trick2019}. $V_{\phi}$ vs $V_R$ is the space in which the `Sausage shape' was first noted by \citet{Belokurov2018}. A Sausage-like structure is clearly visible in the present sample as well. The $\sqrt{V_R^2 + V_Z^2}$ vs $V_{\phi}$ space is similar to a traditional Toomre diagram and shows an overdensity around $V_{\phi}=0$, where \citet{Helmi2018} indicate the \GES structure lies. $E_n$ vs $L_z$ has been used by several studies to select kinematic structures \citep[e.g.][]{Helmi2018, Myeong2018}. We note that our $E_n$ values are not the same as other authors, these values are influenced by the choice of MW potential model. If the \citet{McMillan2017} potential is used \citep[as in e.g.][]{Myeong2018}, then the $E_n$ values are consistent with other studies and the shape our $E_n$ vs $L_z$ distribution is the same. Our sample does not include high $E_n$ stars at $L_z < 0$ as are seen in other studies, probably due to the limited volume sampled resulting from the RVS magnitude limit, therefore we cannot investigate possible Sequoia stars. 

We also define a high-velocity sample in order to compare to the two sequences first noted by \citet{Gaia2018b} and characterized using SkyMapper data by \citet{Sahlholdt2019}. As in \citet{Gaia2018b}, we define tangential velocity as
\begin{equation}
V_T = 4.74/\varpi \sqrt{\mu^2_{\alpha*} + \mu^2_{\delta}}
\end{equation}
and select high-velocity stars as those with $V_T > 200 \kms$. The high velocity subsample contains 5000 stars, 4000 of which meet our giant selection criteria.

\begin{figure*}
\includegraphics[clip,width=1.0\hsize,angle=0,trim=2cm 0.5cm 2cm 1.5cm] {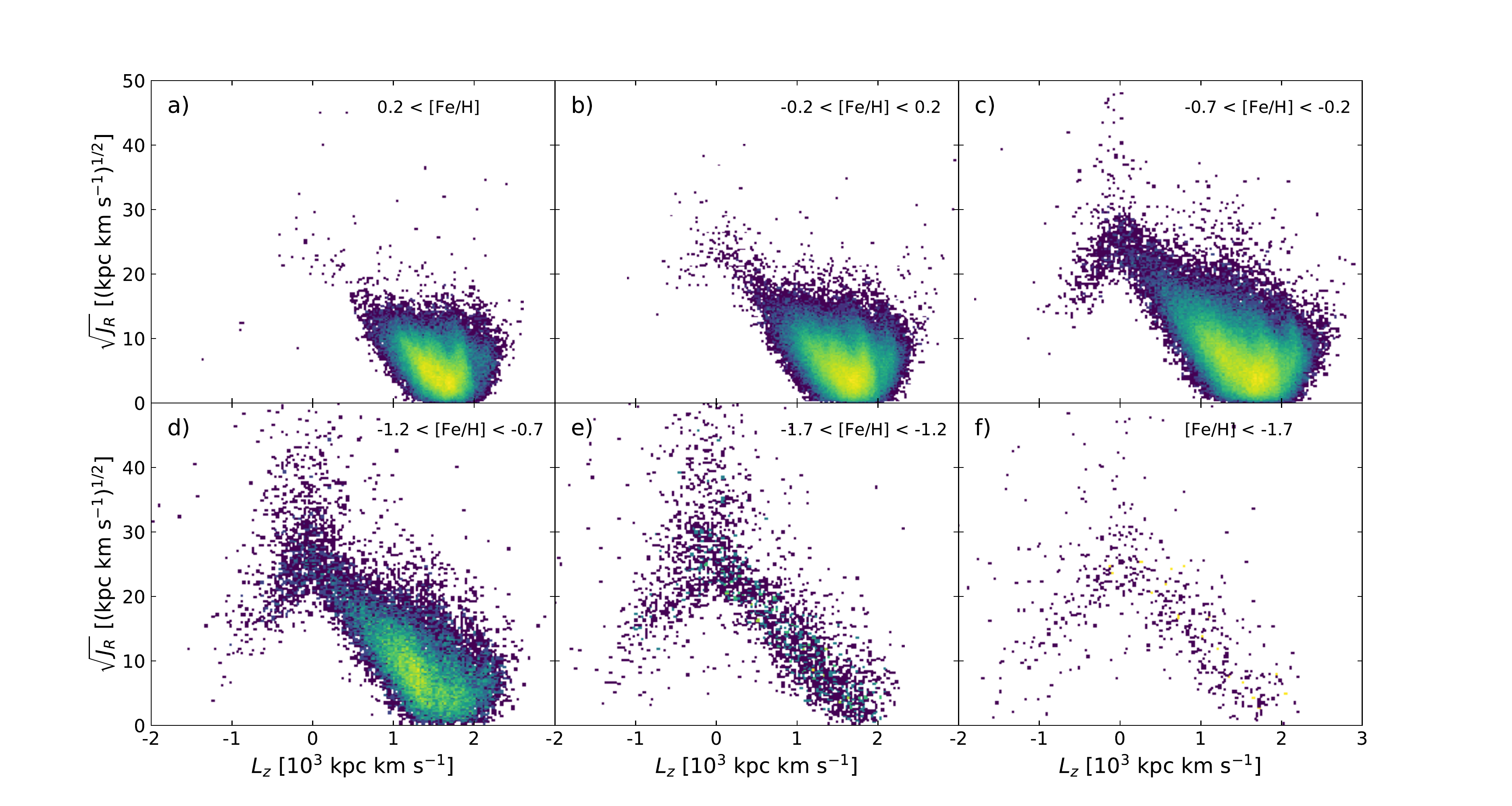}
\caption{$\sqrt{J_R}$ vs $L_z$ action space log(density) distribution for single [Fe/H] subsamples. The [Fe/H] range of each panel is shown in the upper right corner.}
\label{fig:LzJr_feh}
\end{figure*}

To determine the uncertainty in the kinematics of our sample, we preform a Monte Carlo analysis with 10,000 iterations. For each star, the parallax, proper motion, and radial velocity are assigned to a randomly drawn value from a multivariate Gaussian distribution that accounts for the \Gaia correlation between RA proper motion, DEC proper motion, and parallax as well as their individual uncertainties. The uncertainties in the derived kinematics are taken as the standard deviation in the resulting distribution of values. The mean uncertainties are shown in Figure \ref{fig:full_kin} as blue error bars.

We cross match our clean \GES sample, see selection details below, with the APOGEE \citep{Majewski2017} and GALAH \citep{DeSilva2015} surveys, but not enough stars were found to provide significant characterization of the population. 


\section{Analysis}
\label{sec:analysis}

\begin{figure*}
\includegraphics[clip,width=1.0\hsize,angle=0,trim=2cm 0.5cm 2cm 1.5cm] {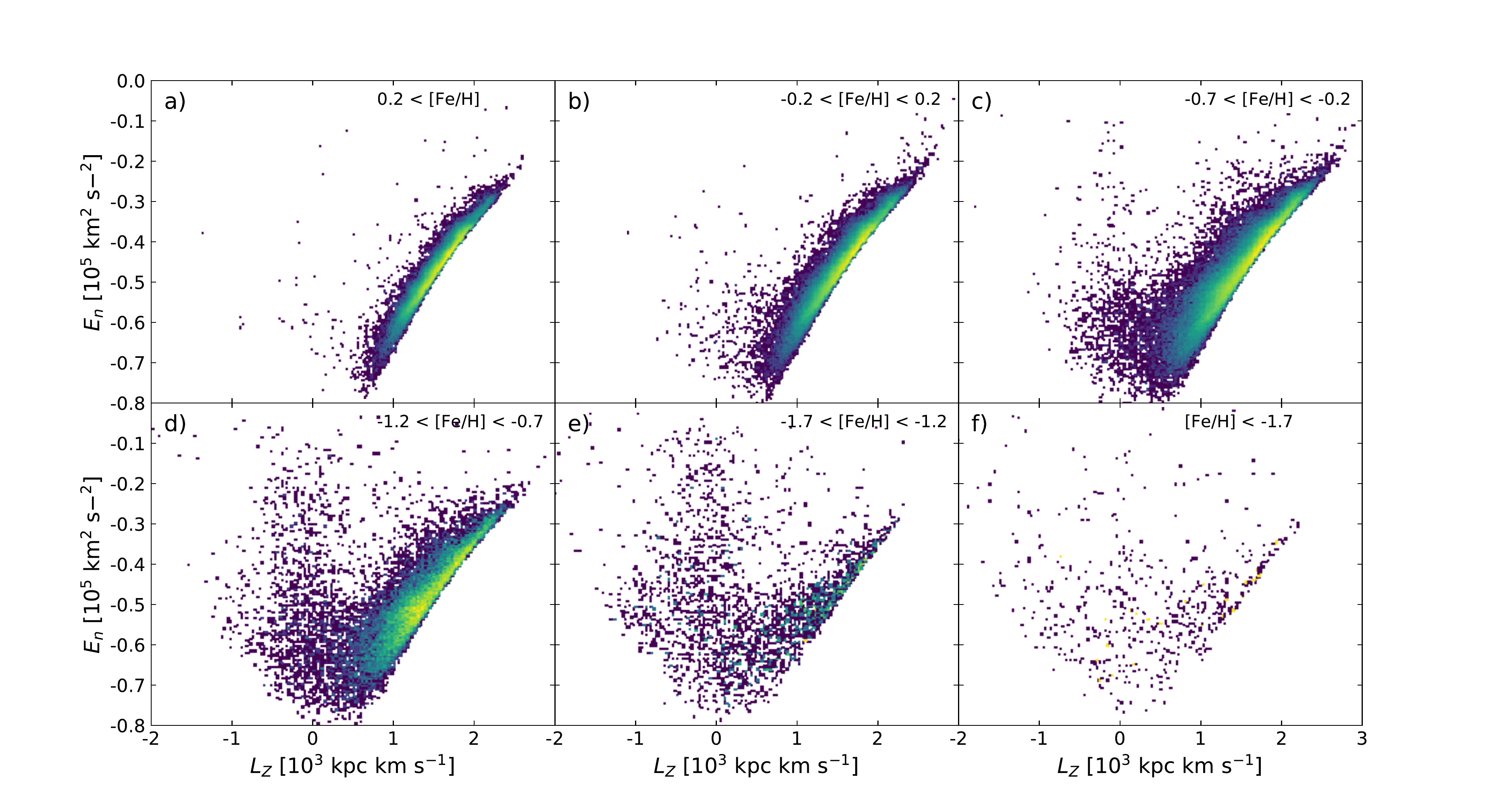}
\caption{$E_n$ vs $L_z$ log(density) distribution for single [Fe/H] subsamples. The [Fe/H] range of each panel is shown in the upper right corner. }
\label{fig:EnLz_feh}
\end{figure*}

In order to explore the SkyMapper-RVS giant sample and search for kinematic features, we inspect the kinematic distributions shown in Figure~\ref{fig:full_kin} binned by [Fe/H]. Figures \ref{fig:LzJr_feh} and \ref{fig:EnLz_feh} show the $\sqrt{J_R}$ vs $L_z$ and $E_n$ vs $L_z$, respectively, for six bins of [Fe/H]. A prominent feature that emerges is the high $J_R$ plume at $L_z \sim 0$ with $-0.7 <$ [Fe/H] $< -1.7$. The plume can also be seen over a range of $E_n$.

We characterize the metallicity of this feature in Figure~\ref{fig:LzJrbins} by inspecting the cumulative metallicity distribution functions (CDFs) of stars binned in $\sqrt{J_R}$ vs $L_z$ space. The bins are selected with the intention of examining the symmetry around $L_z = 0$ and the contamination of the disc. Panel a shows the positions of the bins in $\sqrt{J_R}$ vs $L_z$ space, while panels b -- e show the CDFs of bins with the same range of $J_R$. We performed this analysis using different $L_z$ limits with almost identical results. From these metallicity CDFs we select a `clean \GES' sample that is likely to contain minimal disc contamination. A similar analysis is done in $V_{\phi}$ vs $V_R$ velocity space, Figure~\ref{fig:VphiVrbins}, to characterize the [Fe/H] variations across the Sausage-like feature \citep{Belokurov2018}. Panel a shows bin placement and panels b -- g show the CDFs and bins with the same $V_R$. For reference, the same analysis is done in $\sqrt{V_R^2 + V_Z^2}$ vs $V_{\phi}$ and $E_n$ vs $L_z$ space, shown in Figures \ref{fig:VrzVphibins} and \ref{fig:EnLzbins}, respectively.

\citet{Helmi2018} introduce \GES as a slightly retrograde structure. To investigate this in our sample, we calculate a running mean $L_z$ for the full SkyMapper-RVS giant sample using a bin of 500 stars, independent of any \GES selection, Figure~\ref{fig:LzJr_mean}. 

The SkyMapper survey does not have any selection limitations that would bias our sample. We do not impose a parallax cut, but the RVS magnitude limit will exclude metal-poor dwarfs from our sample as they are too faint. However, we have limited our analysis to the giants, therefore any metallicity bias should be minimal.


\section{Results \& Exploration}

\subsection{SkyMapper-RVS giants}

\begin{figure*}
\includegraphics[clip,width=0.99\hsize,angle=0,trim=1cm 0cm 0cm 0cm]{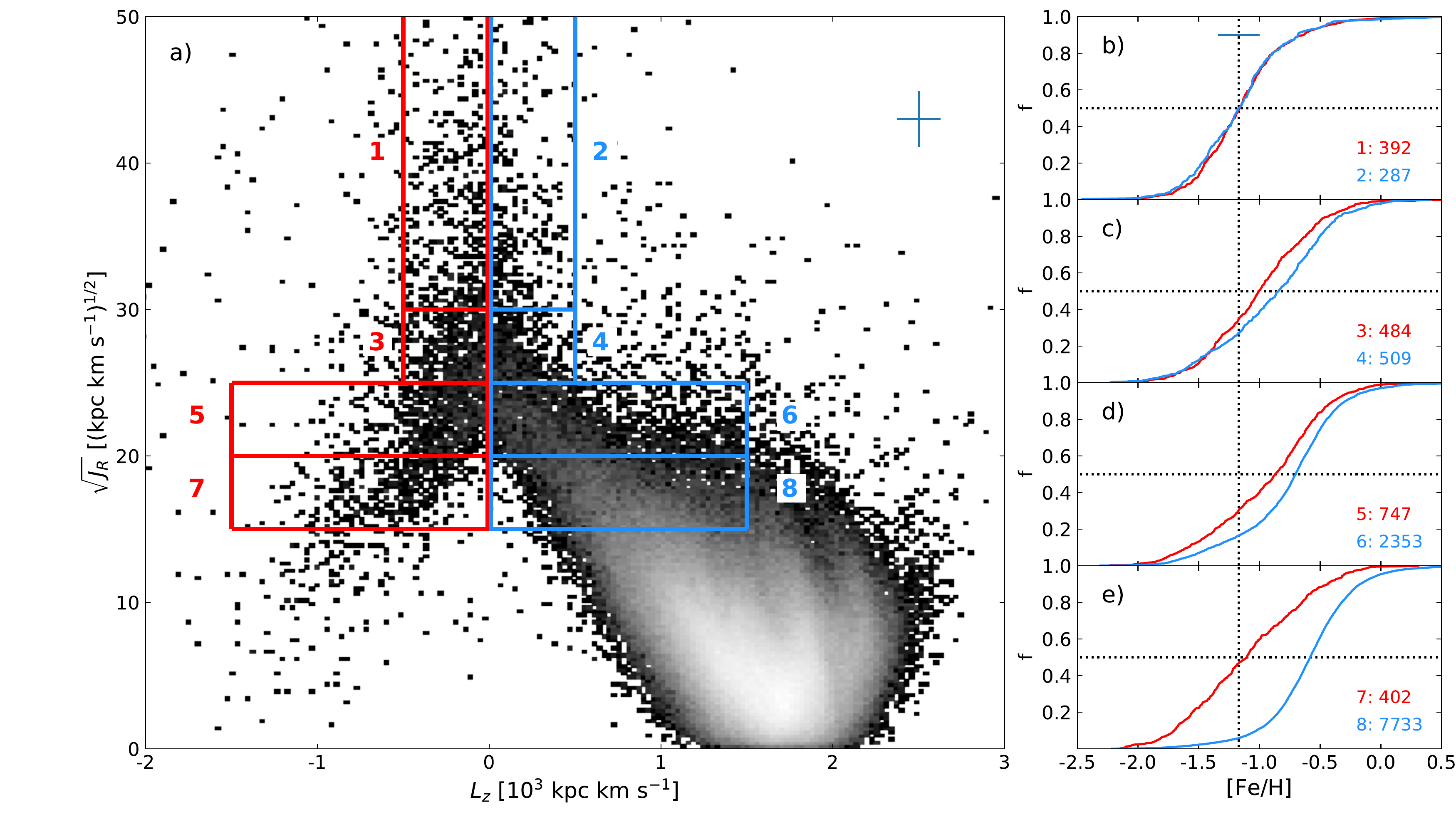}
\caption{The cumulative metallicity distribution functions of bins in $\sqrt{J_R}$ vs $L_z$ action space. The panel a shows the bins positioned to examine symmetry around $L_z=0$. Panels b-e show the CDFs for bins with the same range of $\sqrt{J_R}$. Bins 1 \& 2 are shown in panel b, bins 3 \& 4 are shown in panel c, bins 5 \& 6 are shown in panel d, and bins 7 \& 8 are shown in panel e. The red line shows the retrograde stars and the blue line shows the prograde stars. The bin and number of stars within the bin are indicated. The dotted lines indicate an [Fe/H] $= -1.17$, the median [Fe/H] of bins 1 and 2, and the 50th percentile. The blue error bars represent the mean uncertainty.} 
\label{fig:LzJrbins}
\end{figure*}

The full sample of SkyMapper-RVS giants is composed primarily of disc stars, as can be seen from the kinematics in Figure~\ref{fig:full_kin}. This is expected from the limited volume of the sample. However, there are also a significant number of stars with retrograde motions and non-disc-like orbital properties. We focus our attentions on these non-disc-like stars. We find a large `Sausage-like' structure present in the $V_{\phi}$ vs $V_R$ space (panel b of Figure~\ref{fig:full_kin}) around $V_{\phi} \sim 0$, as first found by \citet{Belokurov2018}. Similarly elongated features with non-rotating kinematics are present in the other panels of Figure~\ref{fig:full_kin}. We therefore explore the characteristics of these potential \GES stars. 

At [Fe/H] below $-1.2$ the kinematically cold disc population is no longer dominant while the retrograde and non-discy stars are present only below [Fe/H] of $-0.2$, see Figures~\ref{fig:LzJr_feh} and~\ref{fig:EnLz_feh}. In particular, the elongated feature around $L_z \sim 0$ covering a large range of $J_R$ and $E_n$ values is strongest at $-0.7 <$ [Fe/H] $< -1.7$, panels c-e. Strongly retrograde stars are all metal-poor, below [Fe/H] $\sim -0.7$, but a low $L_z$ tail is present even at solar [Fe/H]. Stars below [Fe/H] of -1.2 are approximately evenly distributed in $L_z$ suggesting there are either kinematically cold, metal-poor disc stars or halo/accreted stars with disc-like kinematics.

It is interesting that the distribution of our sample looks quite different from that of \citet{Myeong2018}. This is likely because our sample does not extend very far into the halo. We again note that the $E_n$ values are different due to a different choice of MW potential model. 



\subsection{Selecting \GES}
\label{sec:Saus_select}

We start by exploring the [Fe/H] distributions of stars in $L_z$ vs $\sqrt{J_R}$ space in order to characterize the properties of the \GES. 
In Figure~\ref{fig:LzJrbins} we find that at high $J_R$, the retrograde and prograde stars, bins 1 and 2 respectively, have nearly identical CDFs, suggesting that these stars are likely from the same population. The median [Fe/H] of both bins 1 and 2 is $-1.17$. To compare the [Fe/H] distributions of all the bins, [Fe/H] = $-1.17$ and the 50th percentile are indicated by the dotted lines in panels b-e of Figure~\ref{fig:LzJrbins}. The mean uncertainties in $L_z$, $\sqrt{J_R}$, and [Fe/H] are indicated by the blue error bars in panels a and b. 

Bins at $\sqrt{J_R} < 30$ (kpc km s$^{-1})^{1/2}$ are more metal-rich than bins 1 and 2, with the exception of bin 7, and the retrograde stars have different [Fe/H] distributions from the prograde stars. The difference in mean [Fe/H] in all bin, expect bin 7, is roughly greater than or equal to the mean uncertainty in [Fe/H], suggesting a significant difference in [Fe/H] distributions. The prograde stars, in blue, are more metal-rich than the retrograde stars, in red, within a given $J_R$ bin. This reflects the higher fraction of disc stars contained within the prograde bins and is consistent with Figure~\ref{fig:LzJr_feh}. Curiously, the metallicity of the retrograde stars increases with decreasing radial action until bin~7. Bin 7 has a median [Fe/H] similar to bins 1 and 2, but the shape of the CDF is different, indicating a broader metallicity distribution. In panel a of Figure \ref{fig:LzJrbins} one can see that bin 7 contains only truly retrograde stars, as opposed to bins 1, 3, and 5 which contain a large number of stars around $L_z \sim 0$. We suspect that these $L_z \sim 0$ stars are pushing the CDF of those bins towards higher [Fe/H]. The mean [Fe/H] of the retrograde stars with $\sqrt{J_R} < 30$, below bin 7, is $-1.35$, even more metal-poor than bin 7. This is supported by Figure~\ref{fig:LzJr_feh} where one can see the strongly retrograde stars are present only at low metallicity.

The kinematic uncertainties are expected to have a small effect on the resulting CDFs in this analysis. While the height of bins with $\sqrt{J_R} < 30$ is smaller, the number of stars in these bins is larger and the $\sqrt{J_R}$ uncertainties are systematically smaller at lower $\sqrt{J_R}$. The uncertainty in $L_z$ may cause some uncertainty in the comparison of prograde and retrograde stars. However, the $L_z$ uncertainties are small compared to the width of bins 5, 6, 7, and 8, and few stars lie outside bins 1, 2, 3, and 4 in $L_z$. In addition, the resulting CDFs were found to be robust against the exact placement of the bins during analysis.

The strongly retrograde region is where we would expect to find stars belonging to the Sequoia, which \citet{Myeong2019} find to be more metal-poor than the \GES. While this is consistent with the strongly retrograde stars in our sample having lower [Fe/H], due to the limited volume, we do not find any significant signature of the Sequoia. The retrograde metallicity trend and the large range of [Fe/H] in bin 7 suggest that it contains a mix of several populations and the mean [Fe/H] match with bins 1 and 2 is a coincidence. Additionally, the [Fe/H] uncertainties may be too large to distinguish these samples; more precise [Fe/H] measurements may reveal a difference.

\begin{figure*}
\includegraphics[clip,width=0.85\hsize,angle=0,trim=0.5cm 1cm 3.5cm 1cm]{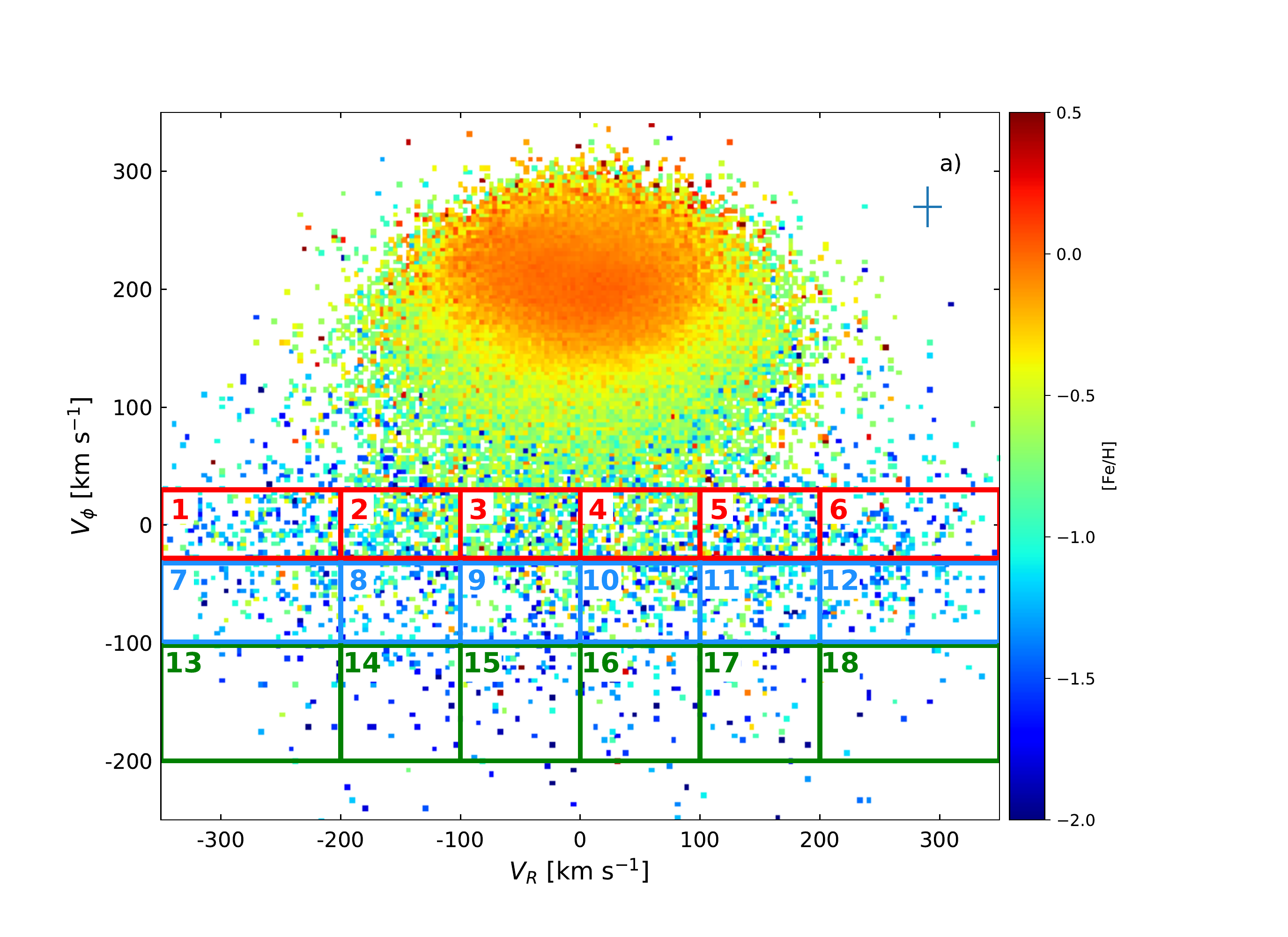}
\includegraphics[clip,width=0.82\hsize,angle=0,trim=0.2cm 0.5cm 0cm 1cm]{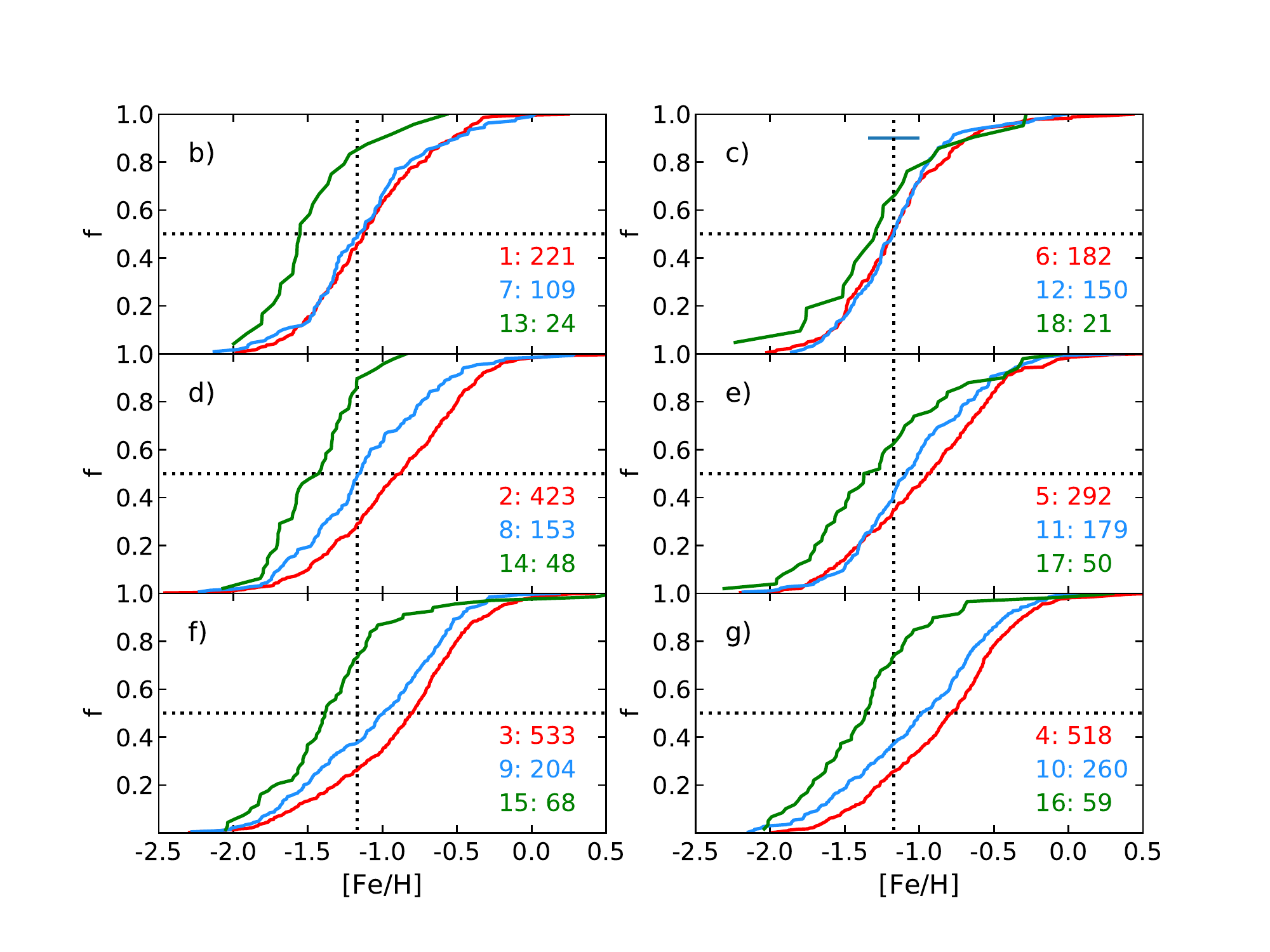}
\caption{The cumulative metallicity distribution functions of bins in $V_{\phi}$ vs $V_R$ velocity space. The top panel (a) shows the distribution of the sample colored by the mean [Fe/H] of each pixel. The bins are positioned to examine the region originally noted by \citet{Belokurov2018} to have a `Sausage-like' shape. The bottom panels show the CDFs of [Fe/H]. Each CDF panel shows bins with the same range of $V_R$. The red line indicates stars with $30 < V_{\phi} < -30$ km s$^{-1}$, the blue line indicates stars with $-30 < V_{\phi} < -100$ km s$^{-1}$, and the green line indicates stars with $-100 < V_{\phi} < -200$ km s$^{-1}$. The bin and number of stars within the bin are indicated. The dotted lines indicate an [Fe/H] $= -1.17$, the median [Fe/H] of bins 1 and 2 in Figure \ref{fig:LzJrbins}, and the 50th percentile. The blue error bars represent the mean uncertainty.} 
\label{fig:VphiVrbins}
\end{figure*}

As \citep{Belokurov2018} initially pointed out the \GES structure in $V_{\phi}$ vs $V_R$ space, we also inspect the [Fe/H] CDFs of bins in that space, Figure~\ref{fig:VphiVrbins}. From panel a one can see that the bins are positioned along the `Sausage-like' structure. Panels b-g show the [Fe/H] CDFs of bins with the same $V_R$. Panels in the same row have the same $|V_R|$. Bins centered around $V_{\phi}=0$ are indicated in red (bins 1-6), bins with small retrograde $V_{\phi}$ are indicated in blue (bins 7-12), and bins with large retrograde $V_{\phi}$ are indicated in green (bins 13-18). As in Figure~\ref{fig:LzJrbins}, the dotted lines indicate [Fe/H] $=-1.17$ and the 50th percentile. 

\begin{figure}
\includegraphics[clip,width=0.99\hsize,angle=0,trim=0.5cm 1cm 2cm 2cm]{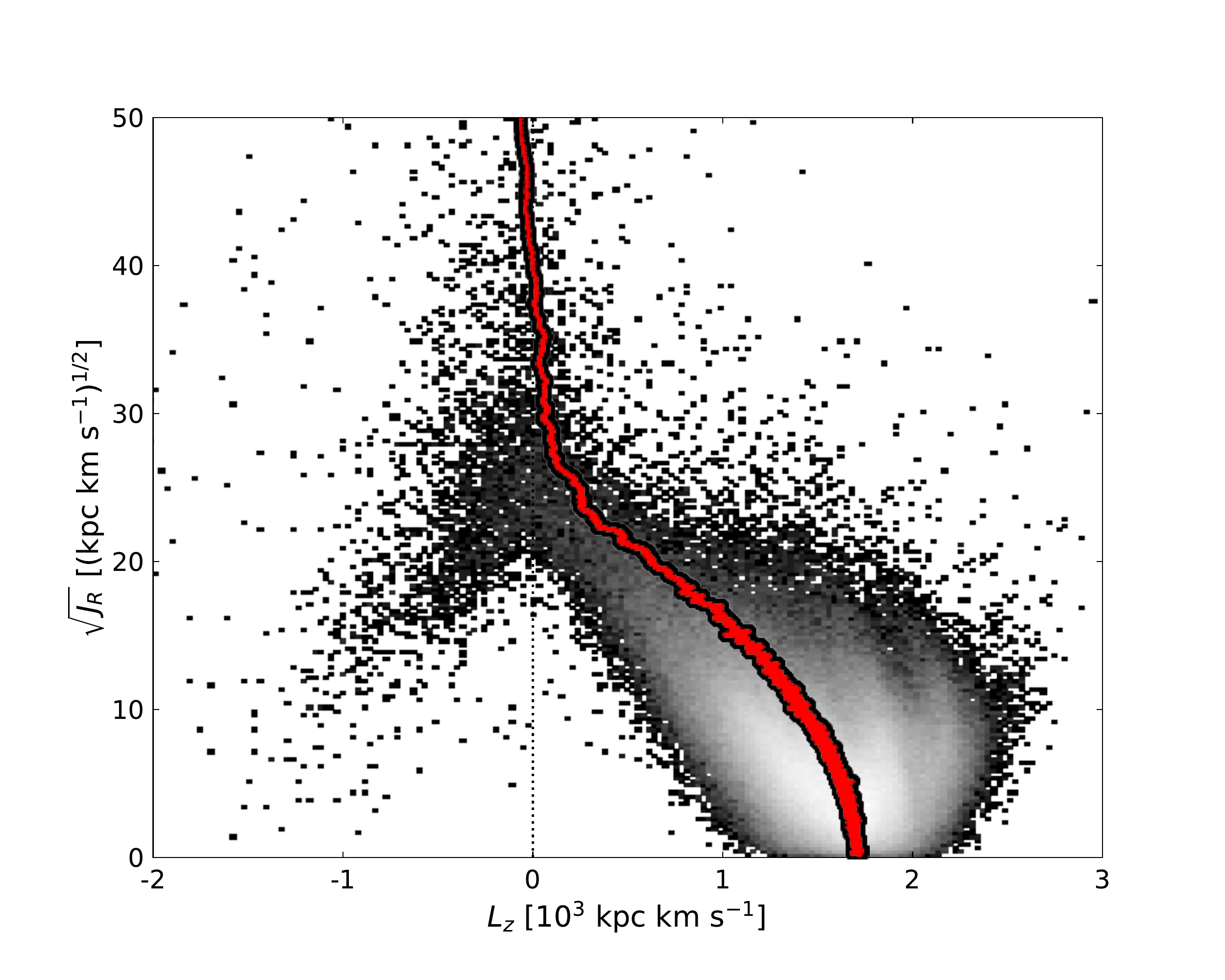}
\caption{$\sqrt{J_R}$ vs $L_z$ action space for the SkyMapper-RVS RGB sample. The red line with black outline shows the running mean $L_z$ using a bin of 500 stars.}
\label{fig:LzJr_mean}
\end{figure}

The green bins with large negative (retrograde) $V_{\phi}$ have small numbers of stars, but are consistently more metal-poor than the other bins even given the [Fe/H] uncertainties. These bins also have median [Fe/H] values that are more metal-poor than any bins in Figure~\ref{fig:LzJrbins} suggesting that in $L_z$ vs $\sqrt{J_R}$ space these metal-poor stars either share a bin with many stars that are more metal-rich or lie outside the binned space.
The median [Fe/H] of the blue bins is equal to or more metal-poor than the red bins. In both red and blue bins, $-100 < V_{\phi} < 30$ km s$^{-1}$, the lower $|V_R|$ bins have a higher median [Fe/H] than the higher $|V_R|$ bins at the same $V_{\phi}$.
The higher resolution of binning in Figure \ref{fig:VphiVrbins} than in Figure \ref{fig:LzJrbins} makes clear the contribution of metal-rich stars at low to retrograde $V_{\phi}$ and low $|V_R|$. 
Bins 1, 6, 7, 8, 11, and 12 have median [Fe/H] values $\sim -1.17$, like bins 1 and 2 in Figure~\ref{fig:LzJrbins}. Bin 18 is also consistent within the [Fe/H] uncertainty, but the low number of stars makes this result fairly uncertain. 

The kinematic uncertainties in $V_{\phi}$ and $|V_R|$ are relatively small compared with the bins. It is possible that these uncertainties are blurring the resulting CDF trends, but the high number of bins makes it likely that the overall trends are robust.

Based on the [Fe/H] CDFs in both $L_z$ vs $\sqrt{J_R}$ and $V_{\phi}$ vs $V_R$ space, we select the stars in bins 1 and 2 of Figure~\ref{fig:LzJrbins} ($L_z$ vs $\sqrt{J_R}$) to be our clean \GES sample, resulting in 679 stars with a median [Fe/H] = $-1.17$.


\subsection{Characterizing \GES}

To characterize the \GES, we use the clean \GES sample, see Section \ref{sec:Saus_select}. We explore the mean $L_z$ of the sample, the [Fe/H] distribution, the color-magnitude diagram (CMD), and where the clean \GES stars lie in other kinematic parameter space. 

Figure~\ref{fig:LzJr_mean} shows the $L_z$ vs $\sqrt{J_R}$ distribution of the giant sample in gray scale with the running mean $L_z$ shown in red. The black outline is present merely to highlight the mean line. We find that the mean $L_z$ decreases with $\sqrt{J_R}$ across the whole $\sqrt{J_R}$ range. 
Above $\sqrt{J_R} \sim 25$ (kpc km s$^{-1})^{1/2}$ the change in mean $L_z$ is very small and $L_z \sim 0$. We also note that stars with high $V_{\phi}$ uncertainties are biased towards retrograde orbits, as demonstrated in figure 3 of \citet{Belokurov2020}. If we limit our analysis to stars with small $V_{\phi}$ uncertainties, the mean $L_z$ of the clean \GES sample increases, but is still slightly retrograde.

Using the clean \GES selection, we can see that bin 1 in Figure~\ref{fig:LzJrbins}, the retrograde bin, has $\sim 100$ more stars in it that bin 2, the prograde bin. Regardless of our \GES selection, visual inspection of Figure \ref{fig:LzJr_feh} shows that the highly radial stars have positive or 0 $L_z$ at [Fe/H] $> -0.7$, mean $L_z \sim 0$ at $-0.7 <$ [Fe/H] $< -1.7$, and negative $L_z$ at [Fe/H] $< -1.7$. Given our MDF of the \GES, it is unlikely that the structure is significantly retrograde. \citet{Helmi2018} conclude that \GES is a slightly retrograde feature, despite a selection that includes significantly retrograde stars ($-1,500 < L_z < 150$ kpc km s$^{-1}$), in contrast to our choice to center our selection on $L_z \sim 0$. We note that the \citet{Koppelman2019} reselection of \GES is centered on $L_Z \sim 0.$

\begin{figure}
\includegraphics[clip,width=0.99\hsize,angle=0,trim=0.5cm 1cm 2cm 2cm]{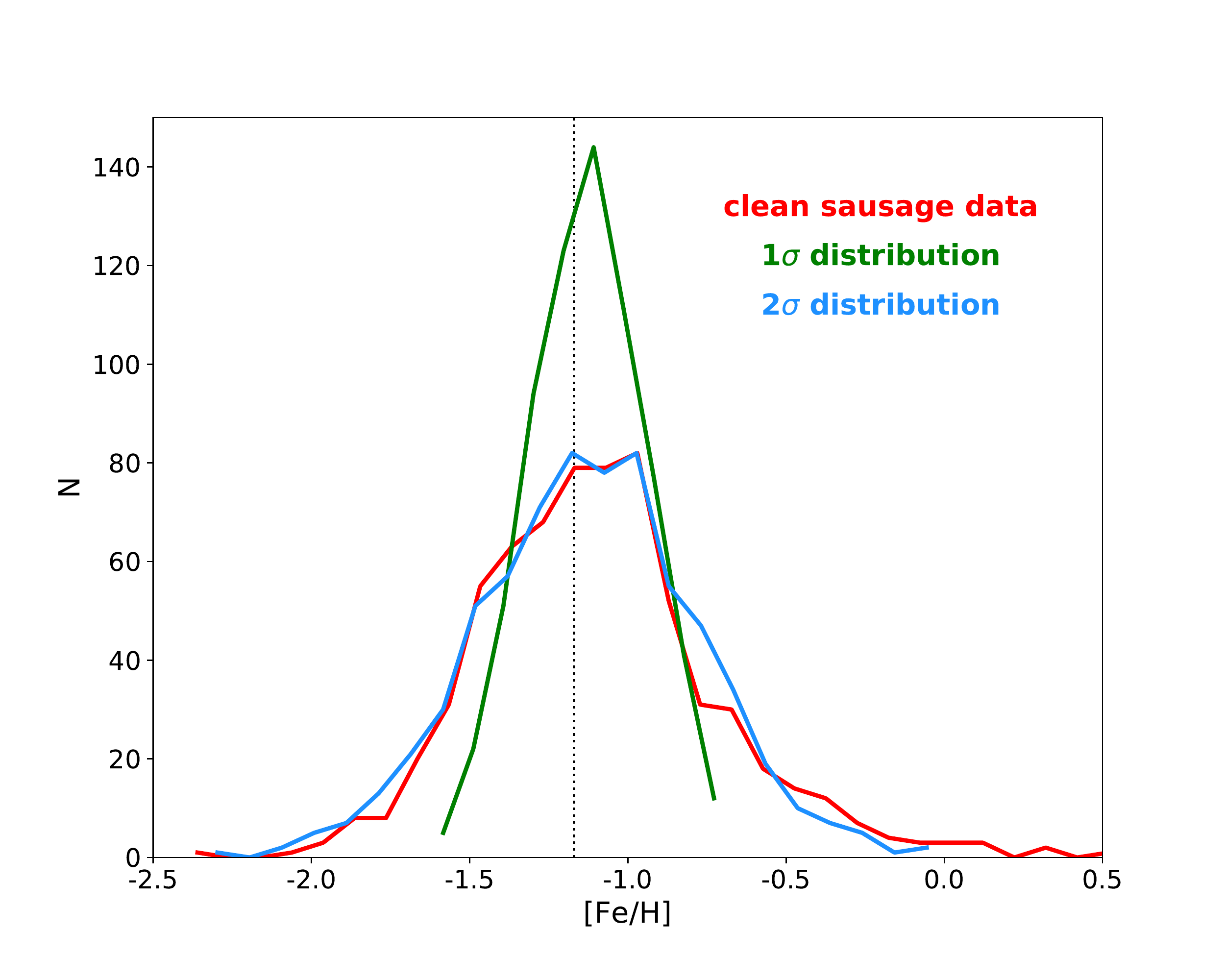}
\caption{The MDF of the clean \GES sample (red) compared to a random Gaussian distribution of the same number of stars with a $\sigma$ of one (green) and two (blue) times the mean [Fe/H] uncertainty of the clean \GES stars. The random Gaussian distributions are centered on the clean \GES 50\% [Fe/H] from Figure~\ref{fig:LzJrbins}, shown as the dotted line.}
\label{fig:MDF_err}
\end{figure}

While the median [Fe/H] of the clean \GES stars is metal-poor, $-1.17$, the [Fe/H] spread of both bin 1 and bin 2 in Figure~\ref{fig:LzJrbins} extends to metal-rich and very metal-poor. In Figure~\ref{fig:MDF_err} we examine the metallicity distribution function (MDF) of the clean \GES stars. The red histogram shows the MDF of the clean \GES stars using the SkyMapper photometric [Fe/H]. The green and blue histograms show a random normal distribution of 680 stars with a mean of $-1.17$ and a dispersion of one and two times the mean [Fe/H] uncertainty of the clean \GES stars, respectively. The $2\sigma$ distribution is in excellent agreement with the clean \GES sample while the $1\sigma$ distribution is too narrow. This suggests that the clean \GES sample is not a single [Fe/H] population, but has a small spread in [Fe/H]. 

\begin{figure}
\includegraphics[clip,width=0.95\hsize,angle=0,trim=0cm 1cm 1cm 2cm]{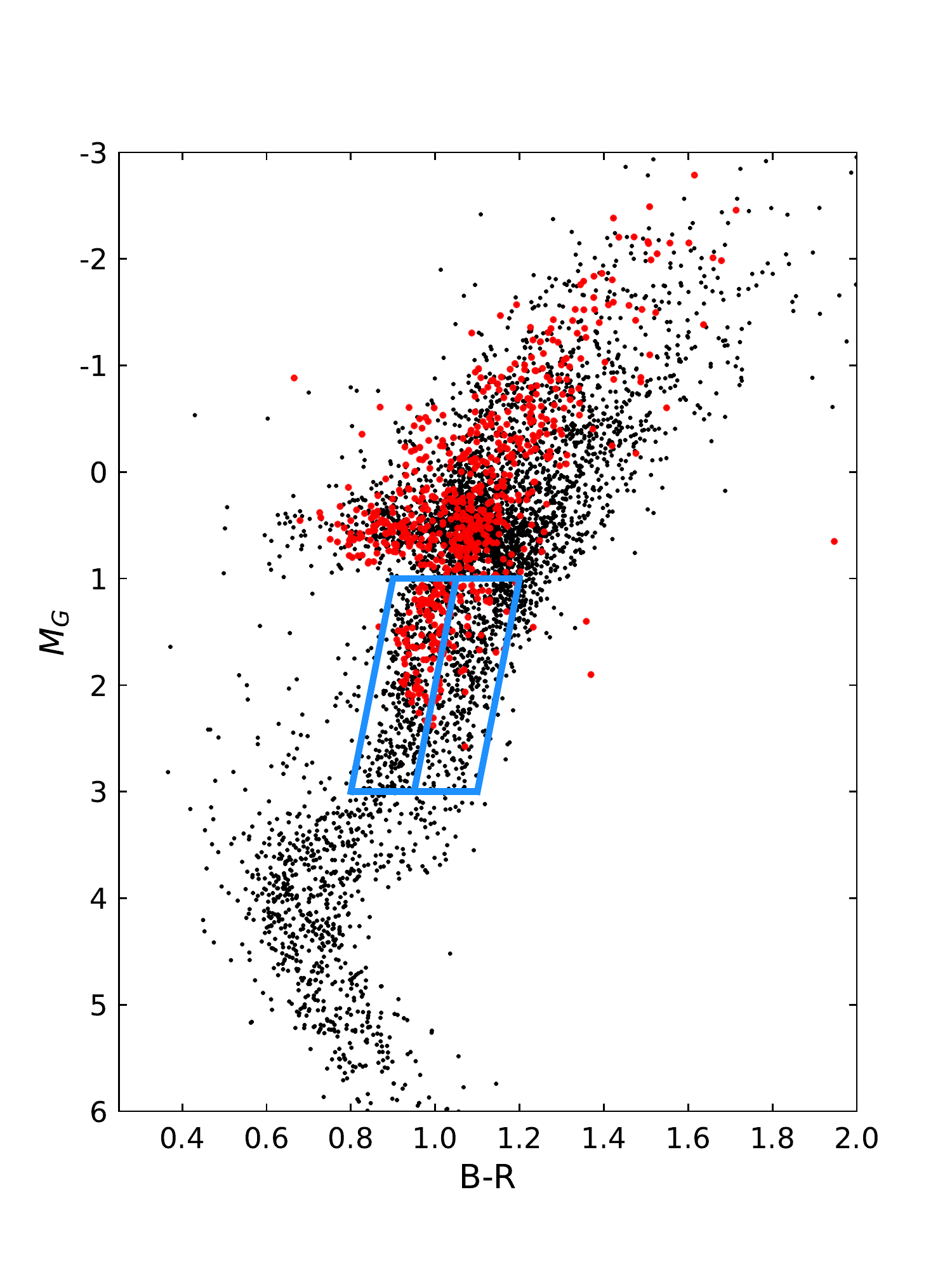}
\caption{The CMD of the high tangential velocity ($V_T$ > 200 km s$^{-1}$) stars in the SkyMapper-RVS sample. The clean \GES sample is shown as red points. The blue boxes indicate the blue and red RGB sequences of \citet{Sahlholdt2019}.}
\label{fig:CMD_saus}
\end{figure}

In Figure~\ref{fig:CMD_saus}, we compare the CMD of the clean \GES sample (red points) with the high $V_T$ stars ($V_T > 200 \kms$) in the SkyMapper-RVS sample (black points). The dual sequence revealed by \Gaia DR2 \citep{Gaia2018b} can be seen in the high $V_T$ stars. The blue boxes show the locations of the two sequences seen by \citet{Sahlholdt2019} in the full SkyMapper sample. The clean \GES stars fall mainly along the blue sequence, although most stars in our sample are farther up the giant branch. We note that the clean \GES stars shown here have not been limited to high $V_T$. 

\begin{figure*}
\includegraphics[clip,width=0.99\hsize,angle=0,trim=2cm 3cm 3cm 3cm] {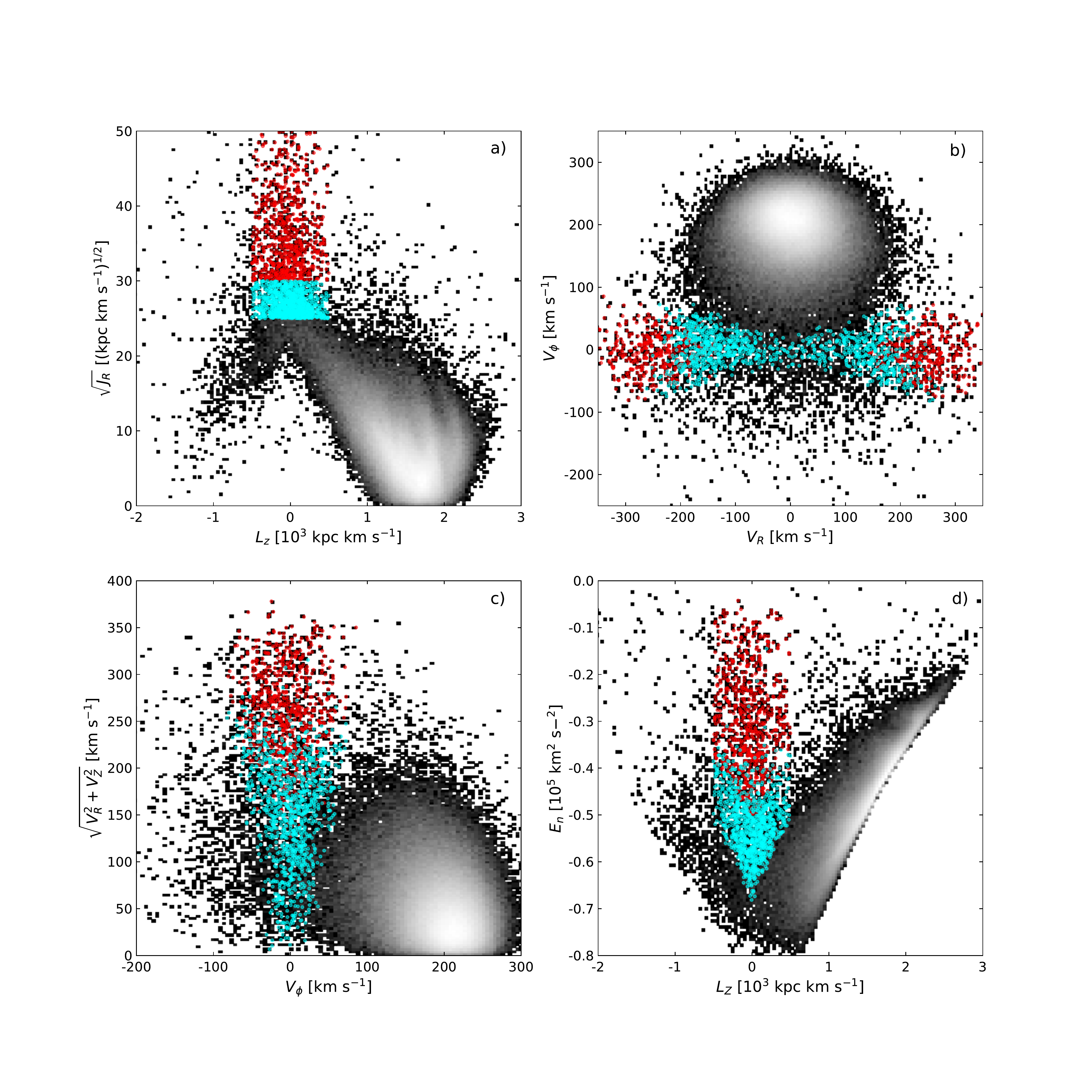}
\caption{Same as Figure~\ref{fig:full_kin} shown is gray scale. In addition, the clean \GES sample is shown as the red points. The cyan points show the stars in bins 3 and 4 from Figure~\ref{fig:LzJrbins}, the sub-\GES stars.}
\label{fig:saus_kin}
\end{figure*}

The mean [Fe/H] of our clean \GES sample is $-1.17$. Estimates of the mean metallicity from previous studies have been more metal-poor; \citet{Helmi2018} find $\sim -1.6$ and \citet{Myeong2019} find $\sim -1.3$. \citet{Sahlholdt2019} find the blue sequence of the high velocity HRD has [Fe/H] $\sim -1.4$ dex, consistent with \citet{Lancaster2019} who find that blue horizontal branch stars in the halo show a metal-rich component peaking at $-1.4$. However, \citet{Amarante2020a} find a mean [Fe/H] of $-1.24$ for the blue sequence using LAMOST metallicities. \citet{Conroy2019} find a mean [Fe/H] of $-1.2$ when selecting highly radial halo stars. We discuss possible reasons for the differences in [Fe/H] in Section \ref{sec:discussion}. 

Full orbits for 10 stars randomly selected from the clean \GES sample were calculated, see Section~\ref{ap:orbits}. These orbits clearly visualize the highly radial nature of the stars, admittedly by design of the selection. They are extremely non-disc-like orbits with perihelions at \Rgal $\leq 2$ kpc and aphelions between 15 and 30 kpc. They vary significantly in z, some staying close to the disc mid-plane, some wandering far from the mid-plane over time, and some orbiting at an angle to the disc.


\subsection{Other Kinematic Spaces}

Although we selected the clean \GES sample in $L_z$ vs $\sqrt{J_R}$ space, most studies have identified these kinematic features in other spaces. Figure~\ref{fig:saus_kin} shows the distribution of the clean \GES stars (red points) in $\sqrt{J_R}$ vs $L_z$ (panel a), $V_{\phi}$ vs $V_R$ (panel b), $\sqrt{V_R^2 + V_Z^2}$ vs $V_{\phi}$ (panel c), and $E_n$ vs $L_z$ (panel d). The full SkyMapper-RVS giant sample is shown in gray scale. For comparison, the stars in bin 3 and bin 4 from Figure~\ref{fig:LzJrbins} are also shown in each panel as cyan points. We will refer to these stars as the sub-\GES sample. From Figure~\ref{fig:saus_kin} it is clear that stars selected to be separate in one kinematic space may overlap significantly in other parameters. 

The distribution of clean \GES stars in $V_{\phi}$ vs $V_R$ (panel b) is consistent with the [Fe/H] CDFs in this space, Figure~\ref{fig:VphiVrbins}, lying mainly in bins 1, 6, 7, and 12. Bins 8 and 11 of Figure~\ref{fig:VphiVrbins} have CDFs consistent with the clean \GES stars, but these regions of $V_{\phi}$ vs $V_R$ space are dominated by the sub-\GES stars. This suggests that the sub-\GES sample contains some true \GES stars. However, the extended distribution of the sub-\GES stars in other kinematic spaces and the corresponding CDFs in Figure~\ref{fig:VphiVrbins} suggest that including the sub-\GES stars would introduce significant contamination into a clean \GES sample. 

While it is possible to imagine defining \GES selection regions in panels a and b of Figure~\ref{fig:saus_kin}, panels c and d appear to be more complex. The `V' shaped distributions of the clean \GES and sub-\GES stars, especially in panel d, illustrate the delicate nature of selecting these kinematic features. The overlap between the clean \GES and sub-\GES stars in $\sqrt{V_R^2 + V_Z^2}$ (panel  c) and $E_n$ (panel d) again suggests that the sub-\GES sample may in fact include some true \GES stars. Likewise the clean \GES sample may have some contamination from non-\GES stars. To visualize the potential contamination of a sample selected in different kinematic spaces, Figure~\ref{fig:slice_kin} shows the distribution of all $\sqrt{J_R}$ bins from Figure~\ref{fig:LzJrbins} in all four kinematic spaces used in this paper.

The distribution of our clean \GES sample is remarkably consistent with the revised selection from \citet{Koppelman2019}, see their figures 3 and 5, in all kinematic spaces. We again note that the $E_n$ values are different due to the choice of Galactic potential model.


\subsection{Weighting \GES}

Next we turn to the question of estimating the mass of the progenitor of the \GES stars.  In this work we find that the merging galaxy has a narrowly defined metallicity distribution function with a clear peak at $-1.2$ dex (see, e.g., Figure~\ref{fig:MDF_err}). There is some intrinsic spread in the distribution but it is small. Given the narrowness of the distribution we can attempt to use the relation between a galaxy's mass and its metallicity to estimate how heavy the \GES progenitor was when it merged with the Milky Way. The mass-metallicity relation is a function of redshift

\begin{figure}
\includegraphics[clip,width=0.99\hsize,angle=0,trim=1cm 0cm 2cm 1cm] {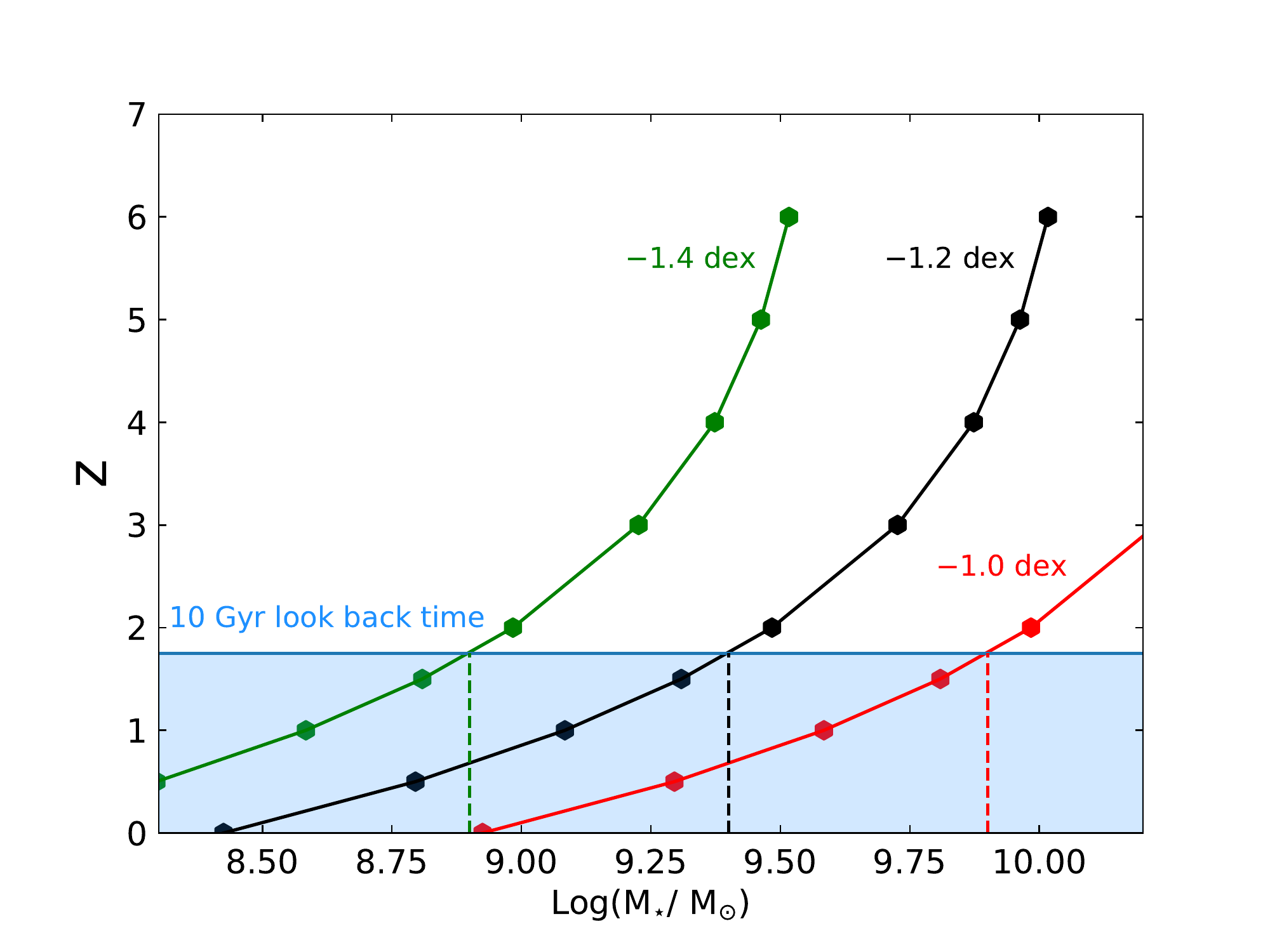}
\caption{A visualization of the mass-metallicity relation as a function of redshift given in Equations~\ref{eq:x1} and \ref{eq:x2} \citep{Ma2016}. Shown is redshift vs. stellar mass for three metallicities ([Fe/H] = $-1.4, -1.2, -1.0$). The blue horizontal line indicates 10 Gyr or $z=1.75$. The dashed vertical lines indicate the stellar mass of the satellite predicted given each metallicity and a peak star forming redshift of 1.75 (10 Gyr look back time).}
\label{fig:mass_metal}
\end{figure}

To derive the (stellar) mass of the \GES requires us to have a rough idea about when the merger happened (and assume that all stars we now see in this structure had formed in the merging galaxy prior to the merger). Several studies have addressed the age of the stars in \GES and/or the high velocity stars in the Milky Way halo as seen in Gaia DR2 \citep[e.g.][]{Gaia2018b, Sahlholdt2019, Gallart2019}. From Figure~\ref{fig:CMD_saus} we may infer that the \GES is essentially associated with the blue sequence in the high-velocity CMD from Gaia DR2 \citep{Gaia2018b}. Turning to \citet{Sahlholdt2019} we note that they find that this particular sequence is all old, in fact all stars can be fit by stellar isochrones of 10 Gyr or older. Thus it appears safe to assume that the merger did not happen earlier than 10 Gyr ago. \citet{Gallart2019} make this conclusion specifically from the cutoff of the blue sequence age distribution and a peak in thick disc stars immediately younger than 10 Gyr.

\citet{Ma2016} studied the evolution of the galaxy mass-metallicity relation. In their Section 3.2 they provide two formulas that quantify the mass-metallicity relation as a function of redshift. 

\begin{equation}
\label{eq:x1}
\log (Z_{\star}/Z_{\odot}) = [\mbox{Fe/H}] + 0.2 = \gamma_{\star} [\log(M_{\star}/M_{\odot}) - 10] + Z_{\star , 10}
\end{equation}

\begin{equation}
\label{eq:x2}
Z_{\star , 10} = 0.67 exp(-0.50z) - 1.04
\end{equation}

These relations should be valid for the range of masses and redshifts of interest to us. \citet{Ma2016} note that their relations do not capture the relations well  for stellar masses above $\sim 10^{11}$\,M$_{\odot}$ at $z < 1$. All indications so far is that the merger happened before $z\sim 1$ and that the mass of the merging galaxy is less than $\sim 10^{11}$\,M$_{\odot}$. It thus appears safe to use these relations to derive the mass of the merging galaxy at the time of the merger. 

Combining Equations~\ref{eq:x1} and \ref{eq:x2} and inverting the equation we get a relation between redshift and stellar mass for a given iron abundance. Figure\,\ref{fig:mass_metal} shows this relation for [Fe/H]$= -1.4, -1.2, -1.0$. Assuming that the merger takes place no later than 10 Gyr ago then our merging galaxy can not be heavier than $\sim 10^{9.4}$\,M$_{\odot}$. Given that these relations are not exact and that the [Fe/H] used by us and by Ma et al. (2016) are not necessarily on exactly the same scale, it appears safe to give the range of possible masses as $10^{8.85} - 10^{9.85}$\,M$_{\odot}$ (i.e. by varying [Fe/H] by $\pm$0.2\,dex).

Our independent stellar mass estimate based on the metallicity relation obtained for a large number of high probability \GES members is well aligned with other estimates in the literature, which find that $\log(M_{\star}/M_{\odot})$ is in the range of 9 to 10. We discuss this result in the context of some of the most recent studies in Section \ref{sec:discussion}.


\section{Discussion and Conclusions}
\label{sec:discussion}

Using a sample of 900,000 stars cross-matched between SkyMapper and \Gaia DR2 RVS, we define a selection of \GES stars in action space that minimizes contamination from Milky Way disc (or other accreted) stars based on the MDFs. Our clean \GES selection results in a sample of 679 stars that were likely accreted during the merger event of \GES. We find that the \GES stars are fairly centered around $L_z = 0$, with a slightly retrograde bias, and are on highly radial orbits. These kinematics, while more constrained, are consistent with previous studies of this structure \citep[e.g.][]{Belokurov2018, Helmi2018, Koppelman2019}.

The real advantage of the current sample is the photometric metallicities that are available from the SkyMapper survey. We find that the \GES stars have a peak [Fe/H] $\sim -1.17$ with a mean uncertainty of 0.17~dex and a relatively small [Fe/H] spread, consistent with a $2 \sigma_{[\mbox{\footnotesize Fe/H}]}$ Gaussian distribution. This consistent with the [Fe/H] distribution of halo stars in highly radial orbits found by the H3 Survey \citep{Conroy2019}. When comparing to the dual CMD sequences in the high $V_T$ halo we find that the \GES stars lie mainly along the blue sequence, consistent with previous work suggesting the blue sequence is comprised of accreted stars \citep[e.g.][]{Haywood2018}. 

Our [Fe/H] measurement is slightly more metal-rich than previous [Fe/H] estimates of \GES stars and the blue sequence \citep[e.g.][]{Helmi2018, Myeong2019, Gallart2019, Sahlholdt2019}, however, these studies used a broader selection in kinematic or CMD space. Based on our characterization of the MDFs of bins in kinematic space, see Figures \ref{fig:LzJrbins}, \ref{fig:VphiVrbins}, \ref{fig:VrzVphibins}, and \ref{fig:EnLzbins}, this likely results in samples contaminated by non-\GES stars. As shown by \citet{Myeong2019} and \citet{Koppelman2019}, the \citet{Helmi2018} estimate is likely contaminated by the Sequoia, Thamnos, and possibly other smaller kinematics structures that have since been identified and found to be more metal-poor than the larger \GES structure. Similarly, the selection of the blue sequence in the CMD by \citet{Sahlholdt2019} likely includes stars from other accreted structures besides the \GES. 

Figures \ref{fig:saus_kin} and \ref{fig:slice_kin} show how the regions of kinematic space used in other studies to select stars belonging to \GES contain significant numbers of stars that we show have a different MDF from the clean \GES sample. While it is almost certain that stars accreted from the \GES lie outside our clean selection region, these regions also contain significant numbers of non-\GES stars, making a robust study of the \GES characteristics difficult. This emphasizes the need for detailed chemical abundance and age measurements in order to robustly identify accreted populations in the halo, especially for ancient mergers that are no longer dynamically distinct. 

Using the galaxy mass-metallicity relation of \citet{Ma2016}, we estimate the mass of the \GES progenitor to be between $10^{8.85}$ and $10^{9.85}\, \mbox{M}_{\odot}$. \citet{Helmi2018} find that \GES stars in APOGEE have a large spread in metallicity and argue that this means a longer time of star formation. They find that the stellar mass of the progenitor was about $6 \cdot 10^8$\,M$_{\odot}$. This mass estimate is consistent with our findings from the mass-metallicity relation, although we do not find as large of a spread in [Fe/H]. 

\citet{Kruijssen2019} estimate the mass of \GES from the metallicity and age of its associated globular cluster to be as high as $10^{9}\, \mbox{M}_{\odot}$. However, they revise their mass estimate in \citet{Kruijssen2020} to be $10^{8.43}\, \mbox{M}_{\odot}$ based on the finding of \citet{Bahe2017} that galaxies in the field have systematically lower metallicities than accreting satellites due to stripping of low-metallicity gas and quenching from the central galaxy. Such an effect could bias our mass estimate to be too high. 

\citet{Vincenzo2019} uses the APOGEE sample from \citet{Helmi2018} to explore the chemical evolution of \GES. Their fiducial model arrives at a galaxy with a large spread in metallicity $-1.26 +0.82/-1.06$\,dex, a median age of $12.33 +0.92/-1.26$\,Gyr, a stellar mass of $10^{10}$\,M$_{\odot}$ at infall, and a gas fraction of $\simeq 0.67$. Again, the stellar sample used to infer this model has a much larger spread in metallicity than our clean sample and it also has two peaks in the metallicity distribution function (their Figure\,2b). Nevertheless, the agreement in stellar mass is good, albeit their model appears to predict a somewhat heavier galaxy than our analysis. 

\citet{Grand2020} analyzed twenty two simulations of Milky Way-like galaxies taken from the Auriga simulations is order to study the effects of a \GES  on the formation of a Milky Way-like galaxy \citep[see][for a discussion of how the simulations were selected]{Grand2020}. Analyzing their simulations they find it likely that the \GES and the Splash \citep{Belokurov2020} are intimately connected, the impact of the \GES resulting in the splashing of pre-existing Milky Way stars. The merger also causes a star-burst, partly fueled by the gas the merging galaxy brought with it \citep[see also][]{Gallart2019}. Here, we are mainly interested in the mass of the merger rather than the impact on the disc formation and subsequent evolution. \citet{Helmi2018} found the merger to be a 1:4 ratio merger, whilst \citet{Grand2020} analysis of their simulations infer a much smaller mass-ratio of the merging galaxy and the Milky Way at the time of the merger. They found that the merging galaxy might be as small as 5\% of the mass of the Milky Way at that time. On the other hand, \citet{Amarante2020b} instead find that no merger is needed to account for the presence of the Splash. In a similar analysis, \citet{Mackereth2019} use the EAGLE simulations to constrain the mass of the \GES progenitor to be $10^{8.5}$ and $10^{9}\, \mbox{M}_{\odot}$.

\citet{Deason2019} derived the total mass of the current stellar halo to be $\sim 1.4 \times 10^9$\,M$_{\odot}$. Our estimate of the mass of \GES progenitor at the time of the merger does not contradict this finding. \citet{Mackereth2020} estimate the total mass of the halo to be $1.3 +0.3/-0.2 \times 10^9$\,M$_{\odot}$, of which $\sim 1 \times 10^9$\,M$_{\odot}$ is accreted. This is also consistent with our estimate, although they only attribute $3 \pm 1 \times 10^8$\,M$_{\odot}$ to the \GES event.

From this work we conclude: 

1) Selection of \GES stars using the following criteria results in the least contaminated sample based on the homogeneity of the MDFs: $30 \leq \sqrt{J_R} \leq 50$ (kpc km s$^{-1})^{1/2}$ and $-500 \leq L_z \leq 500$ kpc km s$^{-1}$.

2) The \GES stars have a relatively narrow [Fe/H] distribution centered at $-1.17$.

3) The \GES stars are consistent with the blue sequence of the dual sequence high tangential velocity stars. However, we stress that not all stars in the blue sequence are necessarily from \GES.

4) From the metallicity and likely age of the stars we predict a stellar mass of $10^{8.85} - 10^{9.85} \mbox{M}_{\odot}$ for the progenitor satellite.

For future work investigating the stellar populations that make up the Milky Way halo we stress that the MDF of the \GES (and other populations) depends on the chosen selection criteria. Therefore detailed elemental abundances combined with kinematics, ages, etc for large samples are required for a well-defined separation of the populations.

\section*{Acknowledgements}

We thank the anonymous referee for their helpful suggestions.
D.K.F., S.F. and C.L.S. were supported by the grant  2016-03412 from the Swedish Research Council. LC is the recipient of the ARC Future Fellowship FT160100402. Parts of this research were conducted by the ARC Centre of Excellence ASTRO 3D, through project number CE170100013.

This work has made use of data from the European Space Agency (ESA) mission {\it Gaia} (\url{https://www.cosmos.esa.int/gaia}), processed by the {\it Gaia} Data Processing and Analysis Consortium (DPAC, \url{https://www.cosmos.esa.int/web/gaia/dpac/consortium}). Funding for the DPAC has been provided by national institutions, in particular the institutions participating in the {\it Gaia} Multilateral Agreement.

\bibliographystyle{mnras}
\bibliography{sausage}

\appendix

\section{Orbits}
\label{ap:orbits}

\begin{figure*}
\includegraphics[clip,width=0.9\hsize,angle=0,trim=0cm 2.8cm 1cm 3cm] {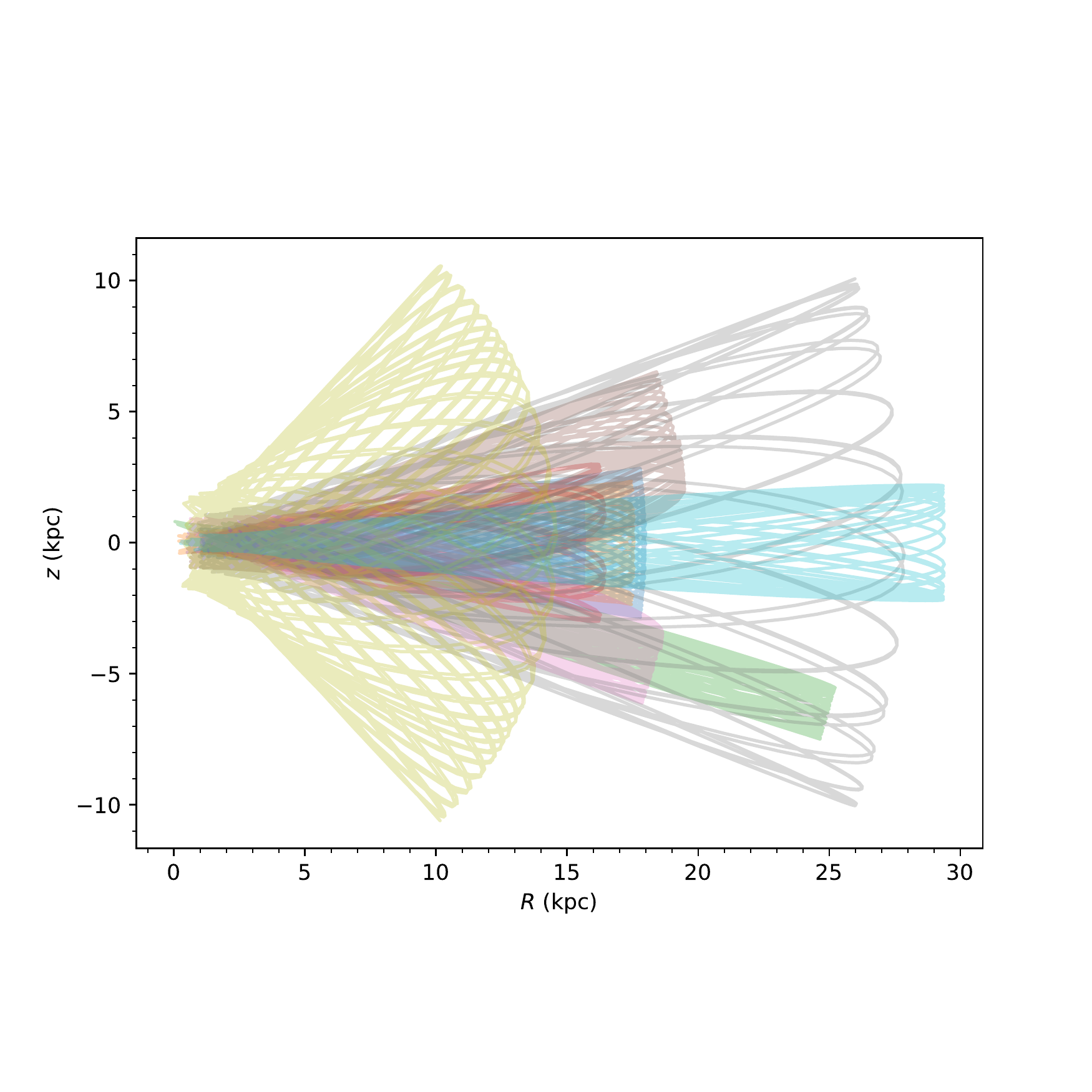}
\caption{Orbits of 10 stars in the clean Sausage sample}
\label{fig:orbits}
\end{figure*}

Full orbits were calculated and visually inspected for $\sim 200$ stars in the clean \GES sample using {\it galpy} with the `MWPotential2014' potential \citep{Bovy2015}. Figure \ref{fig:orbits} shows the orbital path of 10 representative stars in $z$ and $R$. All orbits are highly radial, which is expected from the selection criteria. However, some stars are confined to the disc, while other reach large $z$. 

\section{Additional [Fe/H] CDFs}

\begin{figure*}
\includegraphics[clip,width=0.79\hsize,angle=0,trim=0.5cm 1.1cm 2cm 1.5cm]{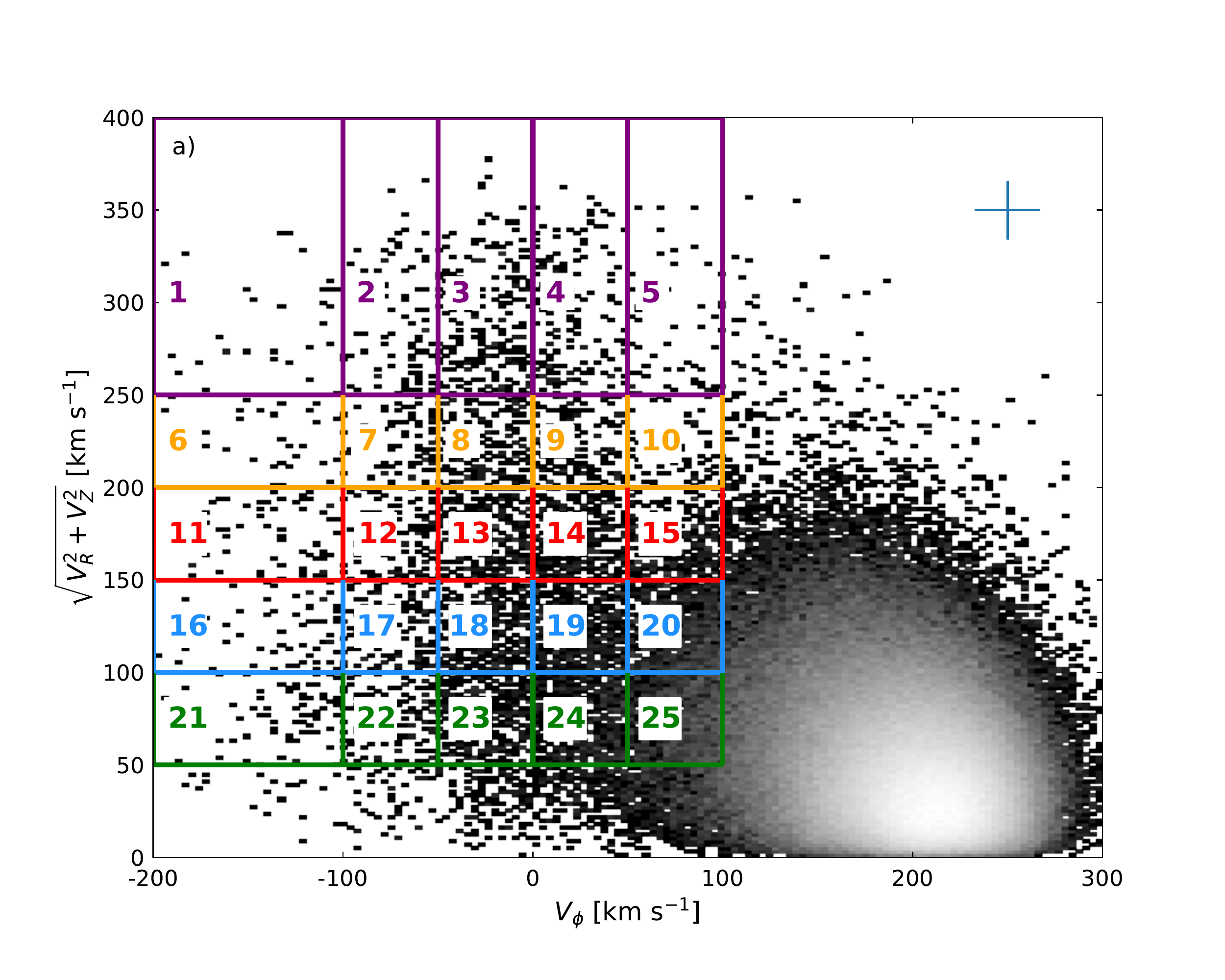}
\includegraphics[clip,width=0.8\hsize,angle=0,trim=0cm 0.7cm 1.5cm 1.3cm]{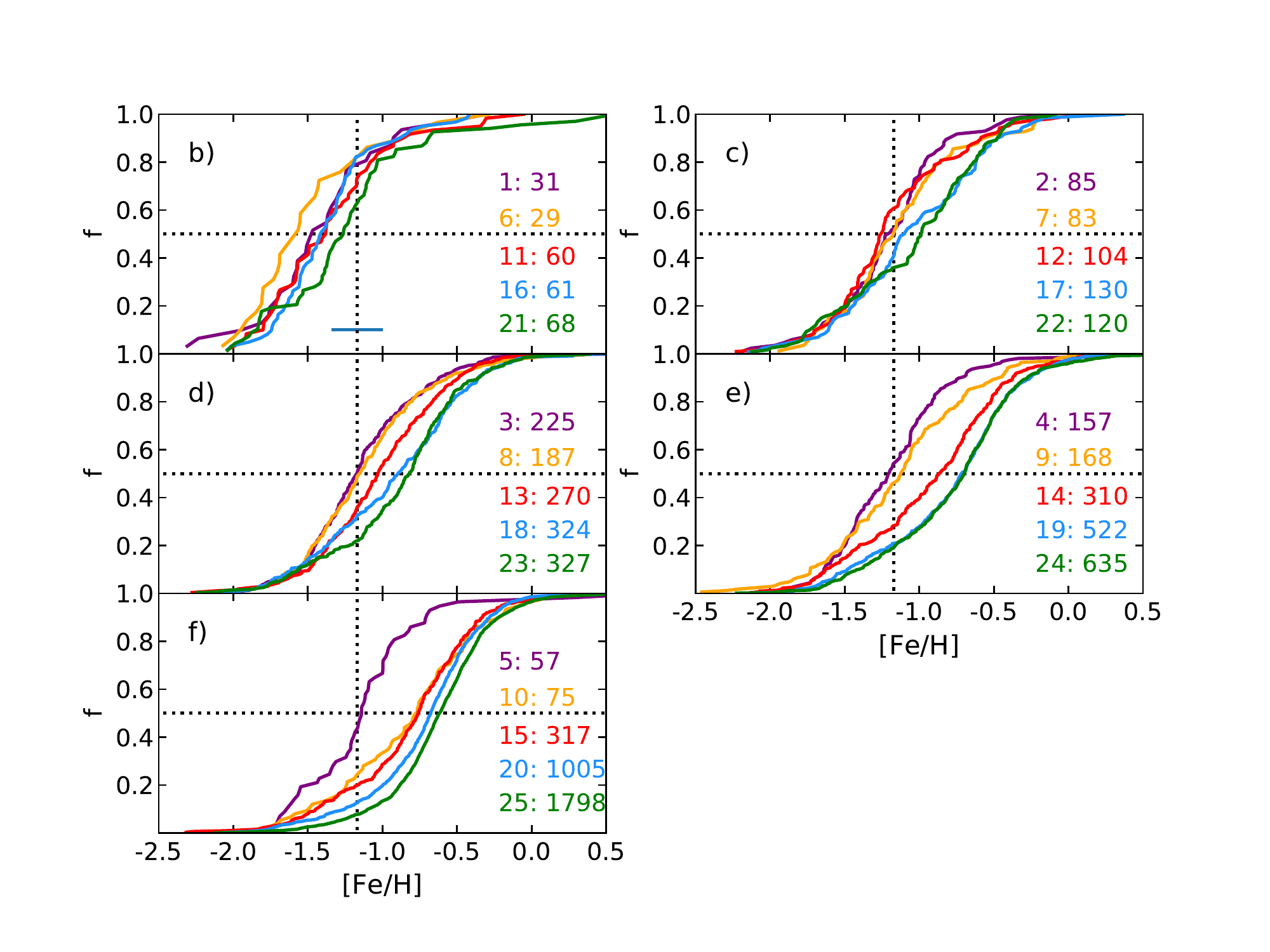}
\caption{The cumulative metallicity distribution functions of bins in $\sqrt{V_R^2 + V_Z^2}$ vs $V_{\phi}$ velocity space. The top panel (a) shows the bins positioned across the space occupied by the {\it Gaia}-Enceladus stars as determined by \citet{Helmi2018}. The bottom panels (b-e) show the CDFs of [Fe/H]. Each CDF panel shows bins with the same range of $V_{\phi}$. The purple lines indicate stars with $250 < \sqrt{V_R^2 + V_Z^2} < 400$ km s$^{-1}$, the yellow lines indicate stars with $200 < \sqrt{V_R^2 + V_Z^2} < 250$ km s$^{-1}$, the red lines indicate stars with $150 < \sqrt{V_R^2 + V_Z^2} < 200$ km s$^{-1}$, the blue lines indicate stars with $100 < \sqrt{V_R^2 + V_Z^2} < 150$ km s$^{-1}$, and the green lines indicate stars with $50 < \sqrt{V_R^2 + V_Z^2} < 100$ km s$^{-1}$. The bin and number of stars within the bin are indicated. The dotted lines indicate an [Fe/H] $= -1.17$, the median [Fe/H] of bins 1 and 2, and the 50th percentile. The blue error bars represent the mean uncertainty.} 
\label{fig:VrzVphibins}
\end{figure*}

\begin{figure*}
\includegraphics[clip,width=0.99\hsize,angle=0,trim=1cm 0cm 0cm 0cm]{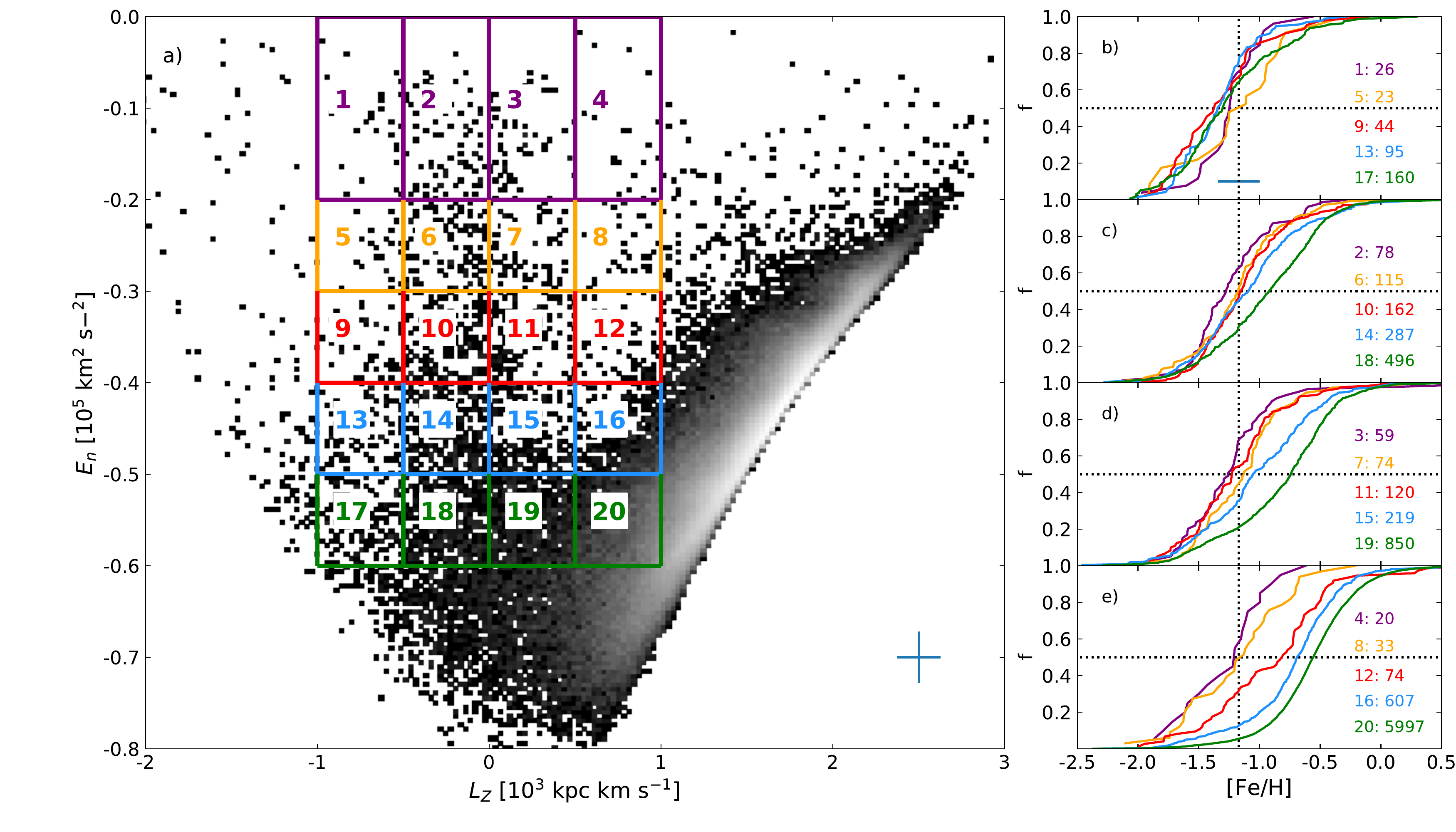}
\caption{The cumulative metallicity distribution functions of bins in $E_n$ vs $L_z$ action space. Panel a shows the bins positioned over the structure occupied by the \GES stars. Panels b-e show the CDFs for bins with the same range of $L_z$. The purple lines indicate stars with $-0.2 < E_n < 0.0$ km$^2$ s$^{-2}$, the yellow lines indicate stars with $-0.3 < E_n < -0.2$ km$^2$ s$^{-2}$, the red lines indicate stars with $-0.4 < E_n < -0.3$ km$^2$ s$^{-2}$, the blue lines indicate stars with $-0.5 < E_n < -0.4$ km$^2$ s$^{-2}$, and the green lines indicate stars with $-0.6 < E_n < -0.5$ km$^2$ s$^{-2}$. The bin and number of stars within the bin are indicated. The dotted lines indicate an [Fe/H] $= -1.17$, the median [Fe/H] of bins 1 and 2, and the 50th percentile. The blue error bars represent the mean uncertainty. }
\label{fig:EnLzbins}
\end{figure*}

Here we show the metallicity CDFs of stars binned in $\sqrt{V_R^2 + V_Z^2}$ vs $V_{\phi}$ velocity space, Figure \ref{fig:VrzVphibins}, and $E_n$ vs $L_z$ action space, Figure \ref{fig:EnLzbins}, similarly to Figures \ref{fig:LzJrbins} and \ref{fig:VphiVrbins}. While we find these do not add significantly to the determination of our \GES selection, they are of interest in characterizing the stars with non-disc kinematics. They also demonstrate the difficulty in selecting a single population of stars based on kinematics alone.

\section{Corresponding kinematic distributions}

Figure~\ref{fig:saus_kin} proved an interesting demonstration of the potential contamination of the \GES population when selected in different kinematic spaces. We provide here a similar figure showing where the stars in all bins of Figure~\ref{fig:LzJrbins} in the other three kinematic spaces examined in the work. We find this figure is a useful tool for connecting the different kinematic and action spaces.

\begin{figure*}
\includegraphics[clip,width=0.99\hsize,angle=0,trim=2cm 4cm 3cm 4cm] {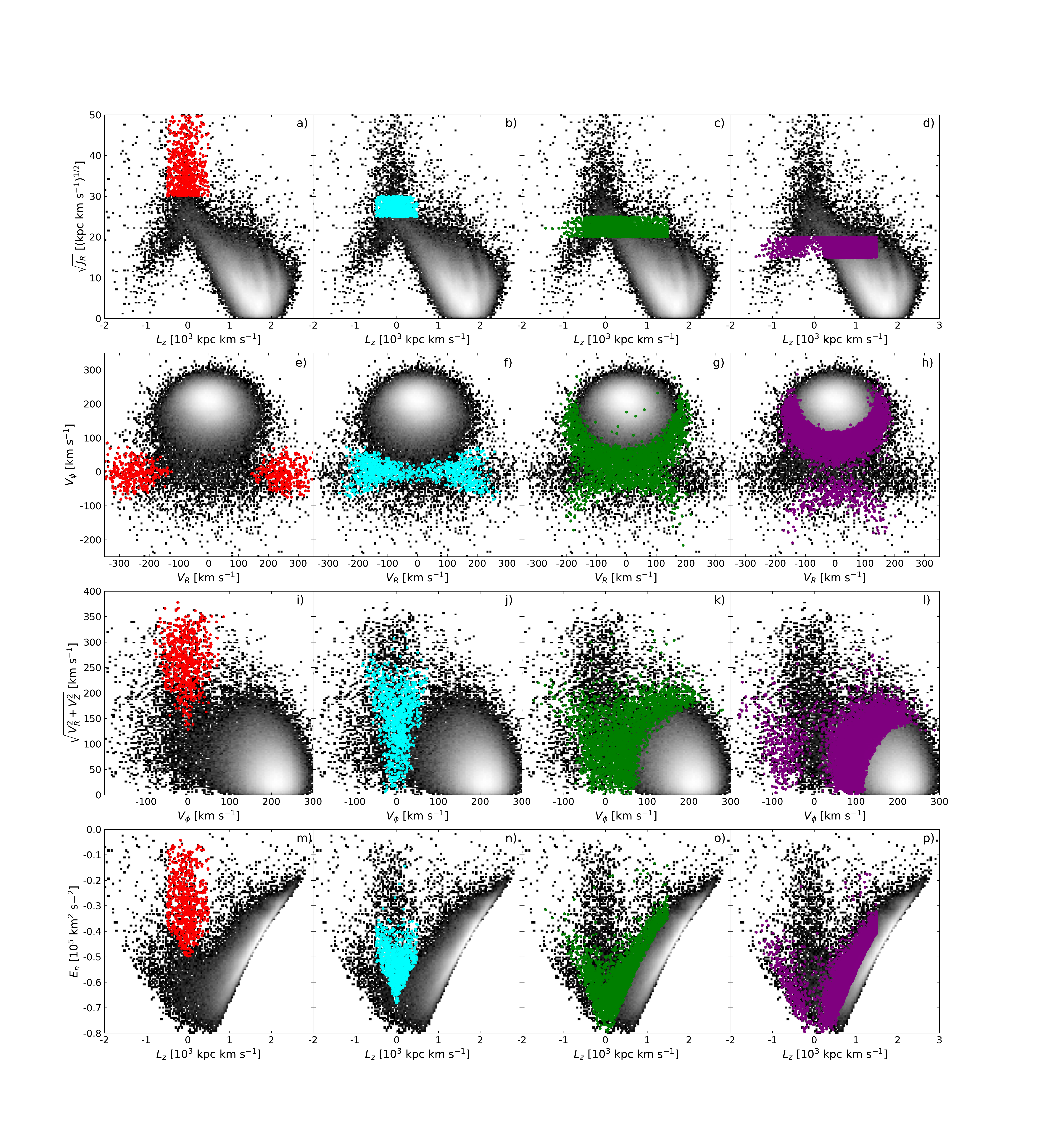}
\caption{The kinematic/action space distributions of stars in `slices' of $\sqrt{J_R}$ corresponding to the $\sqrt{J_R}$ bin pairs in Figure~\ref{fig:LzJrbins}. The gray scale background shows the full SkyMapper-RVS giant sample. The red points shows stars in bins 1 and 2, blue points show stars in bins 3 and 4, green points show stars in bin 5 and 6, and purple points show stars in bins 7 and 8.}
\label{fig:slice_kin}
\end{figure*}

\bsp
\label{lastpage}
\end{document}